# Zipf's law arises naturally in structured, high-dimensional data

Laurence Aitchison[1], Nicola Corradi[2] and Peter E. Latham[1]

[1]Gatsby Computational Neuroscience Unit, UCL, London
[2]Weill Medical College of Cornell University, New York, New York

July 5, 2016

## Abstract

Zipf's law, which states that the probability of an observation is inversely proportional to its rank, has been observed in many domains. While there are models that explain Zipf's law in each of them, those explanations are typically domain specific. Recently, methods from statistical physics were used to show that a fairly broad class of models does provide a general explanation of Zipf's law. This explanation rests on the observation that real world data is often generated from underlying causes, known as latent variables. Those latent variables mix together multiple models that do not obey Zipf's law, giving a model that does. Here we extend that work both theoretically and empirically. Theoretically, we provide a far simpler and more intuitive explanation of Zipf's law, which at the same time considerably extends the class of models to which this explanation can apply. Furthermore, we also give methods for verifying whether this explanation applies to a particular dataset. Empirically, these advances allowed us extend this explanation to important classes of data, including word frequencies (the first domain in which Zipf's law was discovered), data with variable sequence length, and multi-neuron spiking activity.

## 1 Introduction

Both natural and artificial systems often exhibit a surprising degree of statistical regularity. One such regularity is Zipf's law. Originally formulated for word frequency [1], Zipf's law has since been observed in a broad range of domains, including city size [2], firm size [3], mutual fund size [4], amino acid sequences [5], and neural activity [6, 7].

Zipf's law is a relation between rank order and frequency of occurrence: it states that when observations (e.g., words) are ranked by their frequency, the frequency of a particular observation is inversely proportional to its rank,

$$\text{Frequency} \propto \frac{1}{\text{Rank}} \,. \tag{1}$$

Partly because it is so unexpected, a great deal of effort has gone into explaining Zipf's law. So far, almost all explanations are either domain specific or require fine-tuning. For language, there are a variety of domain-specific models, beginning with the suggestion that Zipf's law could be explained by imposing a balance between the effort of the listener and speaker [8–10]. Other explanations include minimizing the number of letters (or phonemes) necessary to communicate a message [11], or by considering the generation of random words [12]. There are also domain-specific models for the distribution of city and firm sizes. These models propose a process in which cities or firms grow by random amounts [2, 3, 13], with a fixed total population or wealth and a fixed minimum size. Other explanations of Zipf's law require fine tuning. For instance, there are many mechanisms that can generate power laws [14], and these can be fine tuned to give an exponent of −1. Possibly the most important fine-tuned proposal is the notion that some systems sit at a highly unusual thermodynamic state — a critical point [6, 15–18].

Only very recently has there been an explanation, by Schwab and colleagues [19], that does not require fine tuning. This explanation exploits the fact that most real-world datasets have hidden structure that can be described using an unobserved variable. For such models — commonly called latent variable models — the unobserved (or latent) variable, $z$, is drawn from a distribution, $P(z)$, and the observation, $x$, is drawn from a conditional distribution, $P(x|z)$. The distribution over $x$ is therefore given by

$$P(x) = \int dz \, P(x|z) \, P(z). \qquad (2)$$

For example, for neural data the latent variable could be the underlying firing rate or the time since stimulus onset.

While Schwab *et al.*'s result was a major advance, it came with some restrictions: the observations, $x$, had to be a high dimensional vector, and the conditional distribution, $P(x|z)$, had to lie in the exponential family with a small number of natural parameters. In addition, the result relied on nontrivial concepts from statistical physics, making it difficult to gain intuition into why latent variable models generally lead to Zipf's law, and, just as importantly, why they sometimes do not. Here we use the same starting point as Schwab *et al.* (Eq. 2), but take a very different theoretical approach — one that is simple and intuitive, and considerably extends our theoretical and empirical understanding of the relationship between latent variable models and Zipf's law. This approach not only gives additional insight into the underlying mechanism by which Zipf's law emerges, but also gives insight into where and how that mechanism breaks down. Moreover, our theoretical approach relaxes the restrictions inherent in Schwab *et al.*'s model [19] (high dimensional observations and an exponential family distribution with a small number of natural parameters). Consequently, we are able to apply our theory to three important types of data, all of which are inaccessible under Schwab *et al.*'s model: word frequencies, models where the latent variable is the sequence length, and complex datasets with high-dimensional observations.

For word frequencies – the domain in which Zipf's law was originally discovered – we show that taking the latent variable to be the part of speech (e.g. noun/verb) can explain Zipf's law. As part of this explanation, we show that if we take only one part of speech (e.g. only nouns) then Zipf's law does not emerge – a phenomenon that is not, to our knowledge, taken into account by any other explanation of Zipf's law for words. For models in which the latent variable is sequence length (i.e. observations in which the dimension of the vector, $x$, is variable), we show that Zipf's law emerges under very mild conditions. Finally, for models that are high dimensional and sufficiently realistic and complex that the conditional distribution, $P(x|z)$, falls outside Schwab *et al.*'s model class, we show that Zipf's law still emerges very naturally, again under mild conditions. In addition, we introduce a quantity that allows us to assess how much a given latent variable contributes to the observation of Zipf's law in a particular dataset. This is important because it allows us to determine, quantitatively, whether a particular latent variable really does contribute significantly to Zipf's law.

## 2 Results

Under Zipf's law, Eq. (1), frequency falls off relatively slowly with rank. This means, loosely, that rare observations are more common than one would typically expect. Consequently, under Zipf's law, one should observe a fairly broad range of frequencies. This is the case, for instance, for words — just look at the previous sentence: there are some very common words (e.g. "a", "that"), and other words that are many orders of magnitude rarer (e.g. "frequencies", "implies"). This is a remarkable property: you might initially expect to see rare words only rarely. However, while a particular rare word (e.g. "frequencies") is far less likely to occur than a particular common word (e.g. "a"), there are far more rare words than common words, and these factors balance almost exactly, so that a random word drawn from a body of text is roughly equally likely to be rare, like "frequencies" as it is to be common, like "a".

Our explanation of Zipf's law consists of two parts. The first part is the above observation — that Zipf's law implies a broad range of frequencies. This observation has been quantified by Mora and Bialek, who showed that a perfectly flat distribution over a range of frequencies is mathematically equivalent to Zipf's law over that range [6] — a result that applies in any and all domains. However, it is important to understand the realistic case: how a finite range of frequencies with an uneven distribution might lead to something similar to, but not exactly, Zipf's law. We therefore extend Mora and Bialek's result, and derive a general relationship that quantifies deviations from Zipf's law for arbitrary distributions over frequency — from very broad to very narrow, and even to multi-modal distributions. That relationship tells us that Zipf's law emerges when the distribution over frequency is sufficiently broad, even if it is not very flat. We complete the explanation of Zipf's law by showing that latent variables can, but do not have to, induce a broad range of frequencies. Finally, we demonstrate theoretically and empirically that, in a variety of important domains, it is indeed latent variables that give rise to a broad range of frequencies, and hence Zipf's law. In particular, we explain Zipf's law in three domains by showing that, in each of them, the existence of a latent variable leads to a broad range of frequencies. Furthermore, we demonstrate that data with both a varying number of dimensions, and fixed but high dimension, leads to Zipf's law under very mild conditions.

## 2.1 A broad range of frequencies implies Zipf's law

By "a broad range of frequencies", we mean the frequency varies by many orders of magnitude, as is the case, for instance, for words: "a" is indeed many orders of magnitude more common than "frequencies". It is therefore convenient to work with the energy, defined by

$$\mathcal{E}(x) \equiv -\log P(x) = -\log \text{Frequency}(x) + \text{const} \tag{3}$$

where, as above, $x$ is an observation, and we have switched from frequency to probability. To translate Zipf's law from observations to energy, we take the log of both sides of Eq. (1) and use Eq. (3) for the energy; this gives us

$$\text{Zipf's law holds exactly} \iff \log r(\mathcal{E}) = \mathcal{E} + \text{const}, \tag{4}$$

where $r(\mathcal{E})$ is the rank of an observation whose energy is $\mathcal{E}$.

Given, as discussed above, that Zipf's law implies a broad range of frequencies, we expect Zipf's law to hold whenever the low and high energies (which translate into high and low frequencies) have about the same probability. Indeed, previous work [6] showed that when the distribution over energy, $P(\mathcal{E})$, is perfectly constant over a broad range, then Zipf's law holds exactly in that range. However, in practice the distribution over energy is never perfectly constant; the real world is simply not that neat. Consequently, to understand Zipf's law in real-world data, it is necessary to understand how deviations from a perfectly flat distribution over energy affect Zipf plots. For that we need to find the exact relationship between the distribution over energy and the rank.

To find this exact relationship, we note, using an approach similar to [6], that if we were to plot rank versus energy, we would see a stepwise increase at the energy of each observation, $x$. Consequently, the gradient of the rank is 0 almost everywhere, and a delta-function at the location of each step,

$$\frac{dr(\mathcal{E})}{d\mathcal{E}} = \sum_x \delta\big(\mathcal{E} - \mathcal{E}(x)\big). \tag{5}$$

The right hand side is closely related to the probability distribution over energy. That distribution can be thought of as a sum of delta-functions, each one located at the energy associated with a particular $x$ and weighted by its probability,

$$P(\mathcal{E}) = \sum_x P(x)\,\delta\big(\mathcal{E} - \mathcal{E}(x)\big) = e^{-\mathcal{E}} \sum_x \delta\big(\mathcal{E} - \mathcal{E}(x)\big), \tag{6}$$

with the second equality following from Eq. (3). This expression says that the probability distribution over energy is proportional to $e^{-\mathcal{E}} \times$ the density of states, a standard result from statistical physics [20]. Comparing Eqs. (5) and Eq. (6), we see that

$$\frac{dr(\mathcal{E})}{d\mathcal{E}} = e^{\mathcal{E}} P(\mathcal{E}). \tag{7}$$

Integrating both sides from $-\infty$ to $\mathcal{E}$ and taking the logarithm gives

$$\log r(\mathcal{E}) = \mathcal{E} + \log P_S(\mathcal{E}) \tag{8}$$

where $P_S(\mathcal{E})$ is $P(\mathcal{E})$ smoothed with an exponential kernel,

$$P_S(\mathcal{E}) \equiv \int_{-\infty}^{\mathcal{E}} d\mathcal{E}' P(\mathcal{E}') e^{\mathcal{E}' - \mathcal{E}}. \tag{9}$$

Comparing Eq. (8) to Eq. (4), we see that for Zipf's law to hold exactly over some range (i.e. $\log r(\mathcal{E}) = \mathcal{E} + \text{const}$, or $r(\mathbf{x}) \propto 1/P(\mathbf{x})$), we need $P_S(\mathcal{E}) = \text{const}$ over that range. This is not new; it was shown previously by Mora and Bialek using essentially the same arguments we used here [6]. What is new is the exact relationship between $P(\mathcal{E})$ and $r(\mathcal{E})$ given in Eq. (8), which is valid whether or not Zipf's law holds exactly. This is important because the distribution over energy is never perfectly flat, so we need to reason about how deviations from $P_S(\mathcal{E}) = \text{const}$ affect Zipf plots — something that our analysis allows us to do. In particular, Eq. (8) tells us that departures from Zipf's law are due solely to variations in $\log P_S(\mathcal{E})$. Consequently, Zipf's law emerges if variations in $\log P_S(\mathcal{E})$ are small compared to the range of observed energies. This requires the distribution over energy to be broad, but not necessarily very flat (see Eq. (22) and surrounding text for an explicit example). Much of the focus of this paper is on showing that latent variable models typically produce sufficient broadening in the distribution over energy for Zipf's law to emerge.

### 2.1.1 Narrow distributions over energy are typical

The analysis in the previous section can be used to tell us why a broad (i.e. Zipfian) distribution over energy is special, and a narrow distribution over energy is generic. Integrating Eq. (6) over a small range (from $\mathcal{E}$ to $\mathcal{E} + \Delta \mathcal{E}$) we see that

$$P(\mathcal{E} \text{ to } \mathcal{E} + \Delta\mathcal{E}) \approx e^{-\mathcal{E}} \mathcal{N}(\mathcal{E} \text{ to } \mathcal{E} + \Delta\mathcal{E}) \tag{10}$$

where $\mathcal{N}(\mathcal{E} \text{ to } \mathcal{E} + \Delta\mathcal{E})$ is the number of states with energy between $\mathcal{E}$ and $\mathcal{E} + \Delta\mathcal{E}$. As we just saw, for a broad, Zipfian distribution over energy, we require $P(\mathcal{E})$ to be nearly constant. Thus, Eq. (10) tells us that for Zipf's law to emerge, we must have $\mathcal{N}(\mathcal{E} \text{ to } \mathcal{E} + \Delta\mathcal{E}) \propto e^{\mathcal{E}}$ (an observation that has been made previously, but couched in terms of entropy rather than density of states [6, 17–19]). However, there is no reason for the number of states to take this particular form, so we do not, in general, see Zipf's law. Moreover, because of the exponential term in Eq. (10), whenever the range of energies is large, even small imbalances between the number of states and the energy lead to highly peaked probabilities. Thus, narrow distributions over energy are generic — a standard result from statistical physics [20].

The fact that broad distributions are not generic tells us that Zipf's law is not generic. However, the above analysis suggests a natural way to induce Zipf's law: stack together many narrow distributions, each with a peak at a different energy. In the following sections we expand on this idea.

## 2.2 Latent variables lead to a broad range of frequencies

We now demonstrate that latent variables can broaden the distribution over energy sufficiently to give Zipf's law. We begin with generic arguments showing that latent variables typically broaden the distribution over energy. We then show empirically that, in three domains of interest, this broadening leads to Zipf's law. We also show that Zipf's law emerges generically in data with varying dimensions and in latent variable models describing data with fixed, but high, dimension.

### 2.2.1 General principles

To obtain Zipf's law, we need a dataset displaying a broad range of frequencies (or energies). It is straightforward to see how latent variables might help: if the energy depends strongly on the latent variable, then mixing across many different settings of the latent variable leads to a broad range of energies. We can formalize this intuition by noting that for a latent variable model, the distribution over $x$ is found by integrating $P(x|z)$ over the latent variable, $z$ (Eq. 2). Likewise, the distribution over energy is found by integrating $P(\mathcal{E}|z)$ over the latent variable,

$$P(\mathcal{E}) = \int dz P(\mathcal{E}|z) P(z). \tag{11}$$

Therefore, mixing multiple narrow (and hence non-Zipfian) distributions, $P(\mathcal{E}|z)$, with sufficiently different means (e.g., coloured lines in Fig. 1A) gives rise to a broad (and hence Zipfian) distribution, $P(\mathcal{E})$ (solid black line Fig. 1A). This tells us something very important: "special" Zipfian distributions, with a broad range of energies, can be constructed merely by combining many "generic" non-Zipfian distributions, each with a narrow range of energies. Critically, to achieve large broadening, the mean energy, and thus the typical frequency, of an observation must depend on the latent variable; i.e. the mean of the conditional distribution, $P(\mathcal{E}|z)$, must depend on $z$. Taking words as an example, one setting of the latent variable should lead mainly to common (and thus low energy) words, like "a", whereas another setting of the latent variable should lead mainly to rare (and thus high energy) words, like "frequencies".

Our mechanism (mixing together many narrow distribution over energy to give a broad distribution) is one of many possible ways that Zipf's law could emerge in real datasets. It is thus important to be able to tell whether Zipf's law in a particular dataset emerges because of our mechanism, or another one. Critically, if our mechanism is operative, even though the full dataset displays Zipf's law (and hence has a broad distribution over energy), the subset of the data associated with any particular setting of the latent variable will be non-Zipfian (and hence have a narrow distribution over energy). In this case, a broad distribution over energy, and hence Zipf's law, emerges because of the mixing of multiple narrow, non-Zipfian distributions (each with a different setting of the latent variable). To complete the explanation of Zipf's law, we only need to explain why, in that particular dataset, it is reasonable for there to be a latent variable that controls the location of the peak in the energy distribution.

Of course there is, in reality, a continuum — there are two contributions to the width of $P(\mathcal{E})$. One, corresponding to our mechanism, comes from changes in the mean of $P(\mathcal{E}|z)$ as the latent variable changes; the other comes from the width of $P(\mathcal{E}|z)$. To quantify the contribution of each mechanism towards an observation of Zipf's law, we use the standard formula for the proportion of explained variance (or $R^2$) to define the proportion of explained energy variance (PEEV; see Methods, Sec. M3 for further details). PEEV gives the proportion of the total energy variance that can be explained by changes in the mean of $P(\mathcal{E}|z)$ as the latent variable, $z$, changes. PEEV ranges from 0, indicating that $z$ explains none of the energy variance, so the latent variable does not contribute to the observation of Zipf's law, to 1, indicating that $z$ explains all of the energy variance, so our mechanism is entirely responsible for the observation of Zipf's law. As an example, we plot energy distributions with a range of values for PEEV (Fig. 1). The black line is $P(\mathcal{E})$, and the coloured lines are $P(\mathcal{E}|z)$ for different settings of $z$. For high values of PEEV, the distributions $P(\mathcal{E}|z)$ are narrow, but have very different means (Fig. 1A). In contrast, for low values of PEEV, the distributions $P(\mathcal{E}|z)$ are broad, yet have very similar means, so the width of $P(\mathcal{E})$ comes mainly from the width of $P(\mathcal{E}|z)$ (Fig. 1C).

## 2.3 Categorical data (word frequencies)

It has been known for many decades that word frequencies obey Zipf's law [1], and many explanations for this finding have been suggested [8–12]. However, none of these explanations accounts for the observation that, while word frequencies overall display Zipf's law (solid black line, Fig. 2B), word frequencies for individual parts of speech (e.g. nouns vs conjunctions) do not (coloured lines, Fig. 2B; except perhaps for verbs, which we discuss below). We can see directly from these plots that the mechanism discussed in the previous section gives rise to Zipf's law: different parts of speech have narrow distributions over energy (coloured lines, Fig. 2A), and they have different means. Mixing across different parts of speech therefore gives a broad range of energies (solid black line, Fig. 2A), and hence Zipf's law. In practice, the fact that different parts of speech have different mean energies implies that some parts of speech (e.g. nouns, like "ream") consist of many different words, each of which is relatively rare, whereas other parts of speech (e.g. conjunctions, like "and") consist of only a few words, each of which is relatively common. We can therefore conclude that Zipf's law for words emerges because there is a latent variable, the part-of-speech, and the latent variable controls the mean energy. We can confirm quantitatively that Zipf's law arises primarily through our mechanism by noting that PEEV is relatively high, 0.58 (for details on how we compute PEEV, see Methods, Sec. M3.1).

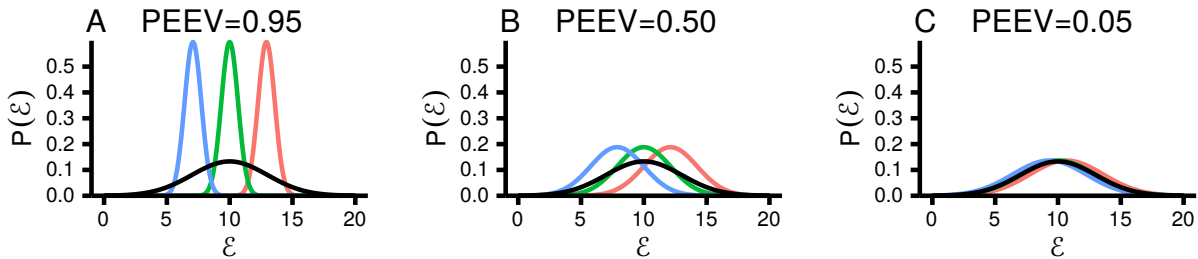

Figure 1: PEEV measures the average width of $P(\mathcal{E}|z)$ relative to $P(\mathcal{E})$. PEEV is close to 0 if the widths are the same, and close to 1 if $P(\mathcal{E}|z)$ is, on average, much narrower that $P(\mathcal{E})$. In all panels, the black line is $P(\mathcal{E})$, and the coloured lines are $P(\mathcal{E}|z)$ for three different settings of the latent variable, $z$. **A**. For high PEEV, the conditional distributions, $P(\mathcal{E}|z)$, are narrow, and have very different means. **B**. For intermediate PEEV, the conditional distributions are broader, and their means are more similar. **C**. For low PEEV, the conditional distributions are very broad, and their means are very similar.

We have demonstrated that Zipf's law for words emerges because of the combination of different parts of speech with different characteristic frequencies. However, to truly explain Zipf's law for words, we have to explain why different parts of speech have such different characteristic frequencies. While this is really a task for linguists, we can speculate. One potential explanation is that different parts of speech have different functions within the sentence. For instance, words with a purely grammatical function (e.g. conjunctions, like "and") are common, because they can be used in a sentence describing anything. In contrast, words denoting something in the world (e.g. nouns, like "ream") are more rare, because they can be used only in the relatively few sentences about that object. Mixing together these two classes of words gives a broad range of frequencies, or energies, and hence, Zipf's law. Finally, using similar arguments, we can see why verbs have a broader range of frequencies than other parts of speech — some verbs (like "is") can be used in almost any context (and one might argue that they have a grammatical function) whereas other verbs (like "gather") refer to a specific type of action, and hence can only be used in a few contexts. In fact, verbs, like words in general, fall into classes [22].

### 2.4 Data with variable dimension

Two models in which the data consists of sequences with variable length have been shown to give rise to Zipf's law [5, 12]. These models fit easily into our framework, as there is a natural latent variable, the sequence length. We show that if the distribution over sequence length is sufficiently broad, Zipf's law emerges.

First, Li [12] noted that randomly generated words with different lengths obey Zipf's law. Here "randomly generated" means the following: a word is generated by randomly selecting a symbol that can be either one of $M$ letters or a space, all with equal probability; the symbols are concatenated; and the word is terminated when a space is encountered. We can turn this into a latent variable model by first drawing the sequence length, $z$, from a distribution, then choosing $z$ letters randomly. Thus, the sequence length, $z$, is "latent", as it is chosen first, before the data are generated — it does not matter that in this particular case, the latent variable can be inferred perfectly from an observation.

Second, Mora *et al.* [5] found that amino acid sequences in the D region of Zebrafish IgM obey Zipf's law.

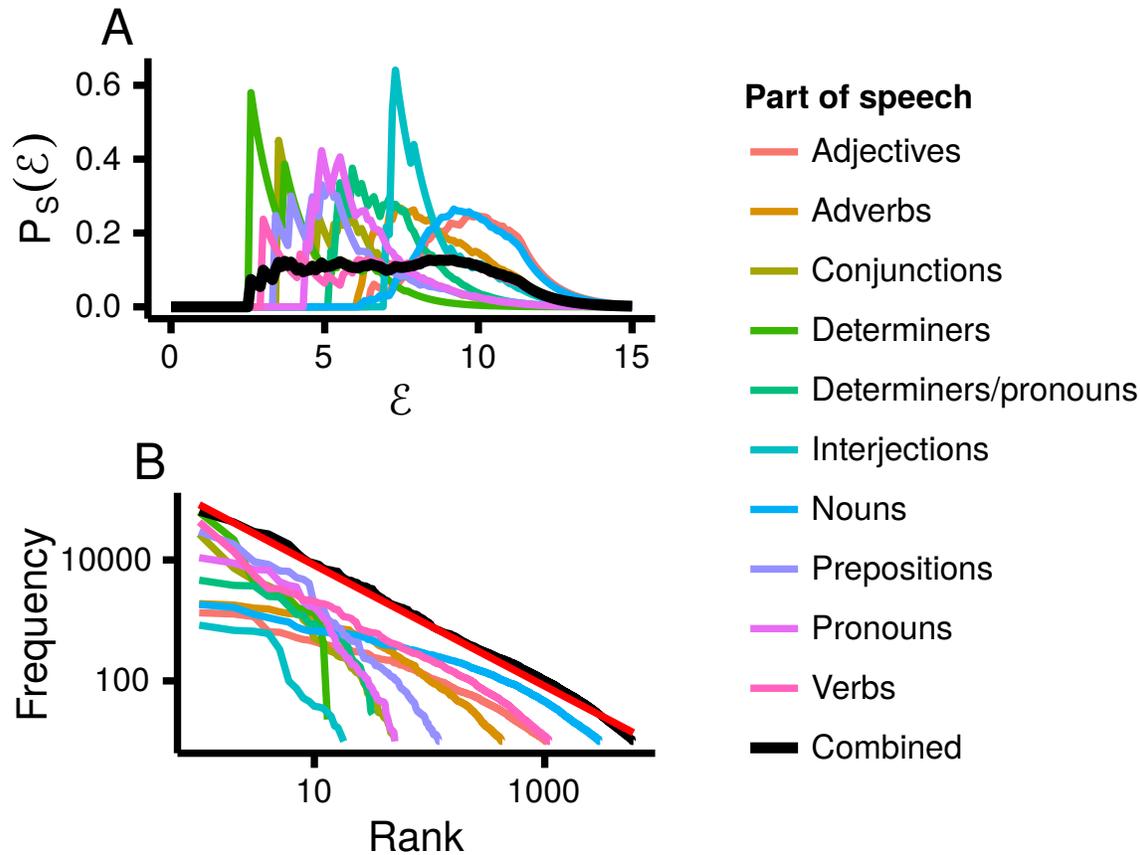

Figure 2: Zipf's law for word frequencies, split by part of speech (data from [21]). The coloured lines are for individual parts of speech, the black line is for all the words. **A**. The distribution over energy is broad for words in general, but the distribution over energy for individual parts of speech is narrow. **B**. Therefore, words in general obey Zipf's law, but individual parts of speech do not (except for verbs, which too can be divided into classes [22]). The red line has a slope of $-1$, and closely matches the combined data.

The latent variable is again $z$, the length of the amino acid sequence. The authors found that, conditioned on length, the data was well fit by an Ising-like model with translation-invariant coupling,

$$P(\mathbf{x}|z) \propto \exp\left(\sum_{i=1}^{z} h(x_i) + \sum_{i,j=1}^{z} J_{|i-j|}(x_i, x_j)\right) \quad (12)$$

where $\mathbf{x}$ denotes a vector, $\mathbf{x} = (x_1, x_2, ..., x_z)$, and $x_i$ represents a single amino acid (of which there were 21).

The basic principle underlying Zipf's law in models with variable sequence length is that there are few short sequences, so each short sequence has a high probability and hence a low energy. In contrast, there are many long sequences, so each long sequence has a low probability and hence a high energy. Mixing together short and long sequences therefore gives a broad distribution over energy and hence Zipf's law.

Models in which sequence length is the latent variable are particularly easy to analyze because there is a simple relationship between the total and conditional distributions,

$$P(\mathbf{x}) = P(z|\mathbf{x}) P(\mathbf{x}) = P(\mathbf{x}|z) P(z). \quad (13)$$

The first equality holds because $z$, the length of the word, is a deterministic function of $\mathbf{x}$, so $P(z|\mathbf{x}) = 1$ (as long as $z$ is the length of the vector $\mathbf{x}$, which is what we assume here); the second follows from Bayes
7

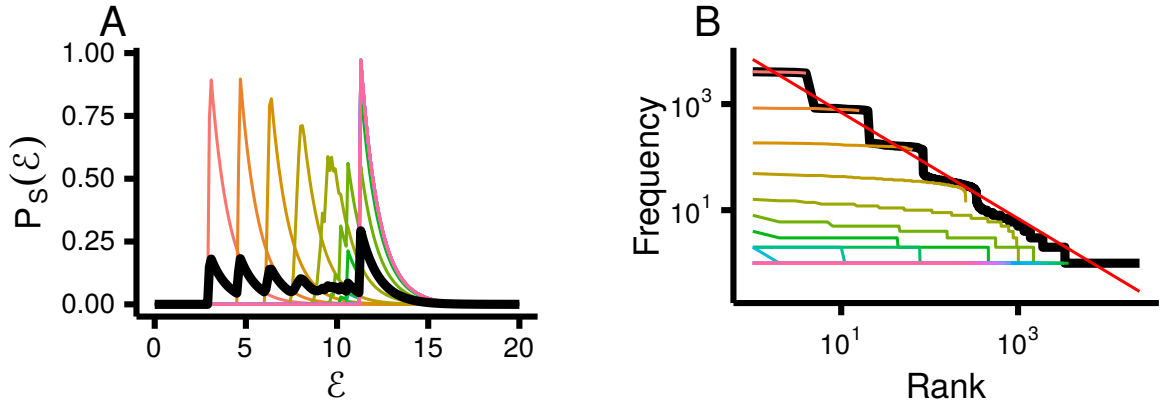

Figure 3: Li's model of random words displays Zipf's law because it mixes words of different lengths. **A**. The distribution over energy. **B**. Zipf plot. In both plots the black lines use all the data and each coloured line corresponds to a different word length. The red line has a slope of $-1$, and so corresponds to Zipf's law.

theorem. To illustrate the general approach, we use this to analyze Li's model (as it is relatively simple). For that model, each element of $\mathbf{x}$ is drawn from a uniform, independent distribution with $M$ elements, so the probability of observing any particular configuration with a sequence length of $z$ is $M^{-z}$. Consequently

$$P(\mathbf{x}) = M^{-z} P(z). \tag{14}$$

Taking the log of both sides of this expression and negating gives us the energy of a particular configuration,

$$\mathcal{E}(\mathbf{x}) = z \log M - \log P(z) \approx z \log M. \tag{15}$$

The approximation holds because $\log P(z)$ varies little with $z$ (in this case its variance cannot be greater than $(M+1)/M$, and in the worst case its variance is $\mathcal{O}((\log \text{Var}[z])^2)$; see Methods, Sec. M4). Therefore, the variance of the energy is approximately proportional to the variance of the sequence length, $z$,

$$\text{Var}[\mathcal{E}(\mathbf{x})] \approx (\log M)^2 \text{Var}[z]. \tag{16}$$

If there is a broad range of sequence lengths (meaning the variance of $z$ is large), then the energy has a broad range, and Zipf's law emerges, as indeed was observed by Li [12]. To confirm that this mechanism gives rise to Zipf's law in Li's model [12], we simulated random words with $M = 4$. As expected, $P(\mathcal{E})$ (black line, Fig. 3A) is relatively flat over a broad range, but the distributions for individual word lengths (coloured lines, Fig. 3A) are extremely narrow. Therefore, data for a single word length does not give Zipf's law (coloured lines, Fig. 3B), but combining across different word lengths does give Zipf's law (black line, Fig. 3B; though with steps, because all words with the same sequence length have the same energy).

Of course, this derivation becomes more complex for models, like the antibody data, in which elements of the sequence are not independently and identically distributed. However, even in such models the basic intuition holds: there are few short sequences, so each short sequence has high probability and low energy, whereas the opposite is true for longer sequences. In fact, the energy is still approximately proportional to sequence length, as it was in Eq. (15), because the number of possible configurations is exponential in the sequence length, and the energy is approximately the logarithm of that number (see Methods, Sec. M5, for a more principled explanation). Consequently, in general a broad range of sequence lengths gives a broad distribution over energy, and hence Zipf's law.

However, as discussed in Sec. 2.2.1 above, just because a latent variable could give rise to Zipf's law does not mean it is entirely responsible for Zipf's law in a particular dataset. To quantify the role of sequence length in Mora *et al.*'s antibody data, we computed PEEV (the proportion of the variance of the energy explained by sequence length) for the 14 datasets used in their analysis. As can be seen in Fig. 4A, PEEV is generally small; less than 0.5 in 12 out of the 14 datasets. And indeed, for the dataset with the smallest PEEV (0.07), Zipf's law is obeyed at each sequence length (Fig. 4B). This in fact turns out to hold for all the datasets, even the one with the highest PEEV (0.72; Fig. 4C).

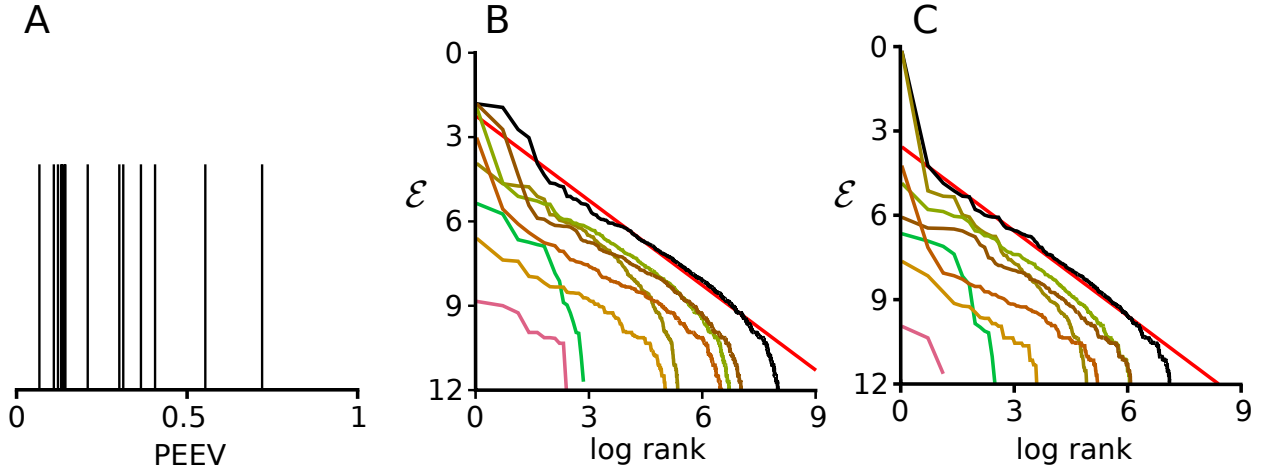

Figure 4: Re-analysis of amino acid sequences in the D region of 14 Zebrafish. **A**. Proportion of the variance explained by sequence length (PEEV) for the 14 datasets. Most are low, and all but two are less than 0.5. **B** and **C**. Zipf plots for the dataset with the lowest (B) and highest (C) PEEV. In both plots the black line uses all the data and the coloured lines correspond to sequence lengths ranging from 1 to 7. The red line has a slope of $-1$, and so corresponds to Zipf's law. Data from Ref. [5], kindly supplied by Thierry Mora. (Note that $\mathcal{E}$ increases downward on the $y$-axis, in keeping with standard conventions.)

The fact that Zipf's law is observed at each sequence length complicates the interpretation of this data. Our mechanism — adding together many distributions, each at different mean energy — plays a role in producing Zipf's law over the whole dataset. However, as indicated in Figs. 4B and C, that's clearly not the whole story; an additional mechanism is needed to explain the fact that Zipf's law emerges at each sequence length. And indeed, such a mechanism has been found: a recent study showed that antibody data is well modeled by random growth and decay processes [23], which leads to Zipf's law at each sequence length.

## 2.5 High-dimensional data

A very important class of models are those where the data is high-dimensional. We show two things for this class. First, the distribution over energy is broadened by latent variables — more specifically, for latent variable models, the variance typically scales as $n^2$. Second, the $n^2$ scaling is sufficiently large that deviations from Zipf's law are negligible in the large $n$ limit.

The reasoning is the same as it was above: we can obtain a broad distribution over energy by mixing together multiple, narrowly peaked (and thus non-Zipfian) distributions. Intuitively, if the peaks of those distributions cover a broad enough range of energies, Zipf's law should emerge. To quantify this intuition, we use the law of total variance [24],

$$\text{Var}_{\mathbf{x}}\left[\mathcal{E}(\mathbf{x})\right] = \text{Var}_z\left[\text{E}_{\mathbf{x}|z}\left[\mathcal{E}(\mathbf{x})\right]\right] + \text{E}_z\left[\text{Var}_{\mathbf{x}|z}\left[\mathcal{E}(\mathbf{x})\right]\right] \tag{17}$$

where again $\mathbf{x}$ is a vector, this time with $n$, rather than $z$, elements. This expression tells us that the variance of the energy (the left hand side) must be greater than the variance of the mean energy (the first term on the right hand side). (As an aside, this decomposition is the essence of PEEV; see Methods, Sec. M3.)

As discussed on Sec. 2.2.1, the reason latent variable models often lead to Zipf's law is that the latent variable typically has a strong effect on the mean energy (see in particular Fig. 1). We thus focus on the first term in Eq. (17), the variance of the mean energy. We show next that it is typically $\mathcal{O}(n^2)$, and that this is sufficiently broad to induce Zipf's law.

The mean energy is given by,

$$\mathrm{E}_{\mathbf{x}|z}\left[\mathcal{E}(\mathbf{x})\right] = -\sum_{\mathbf{x}} P(\mathbf{x}|z) \log P(\mathbf{x}). \tag{18}$$

This is somewhat unfamiliar, but can be converted into a very standard quantity by noting that in the large $n$ limit we may replace $P(\mathbf{x})$ with $P(\mathbf{x}|z)$, which converts the mean energy to the entropy of $P(\mathbf{x}|z)$. To see why, we write

$$\mathrm{E}_{\mathbf{x}|z}\left[\mathcal{E}(\mathbf{x})\right] = -\sum_{\mathbf{x}} P(\mathbf{x}|z) \log P(\mathbf{x}|z) + \sum_{\mathbf{x}} P(\mathbf{x}|z) \log \frac{P(\mathbf{x}|z)}{P(\mathbf{x})}. \tag{19}$$

For low dimensional latent variable models (more specifically, for models in which $z$ is $k$ dimensional with $k \ll n$), the second term on the right hand side is $\mathcal{O}(k/2 \log n)$. Loosely, that's because it's positive and its expectation over $z$ is the mutual information between $\mathbf{x}$ and $z$, which is typically $\mathcal{O}(k/2 \log n)$ [25]. Here, and in almost all of our analysis, we consider low dimensional latent variables; in this regime, the second term on the right hand side is small compared to the energy, which is $\mathcal{O}(n)$ (recall, from the previous section, that the energy is proportional to the sequence length, which here is $n$; see also Methods, Sec. M5). Thus, in the large $n$ and small $k$ limit — the limit of interest — the second term can be ignored, and the mean energy is approximately equal to the entropy of $P(\mathbf{x}|z)$,

$$\mathrm{E}_{\mathbf{x}|z}\left[\mathcal{E}(\mathbf{x})\right] \approx -\sum_{\mathbf{x}} P(\mathbf{x}|z) \log P(\mathbf{x}|z) \equiv H_{\mathbf{x}|z}(z). \tag{20}$$

Approximating the energy by the entropy is convenient because the latter is intuitive, and often easy to estimate. This approximation breaks down (as does the $\mathcal{O}(k/2 \log n)$ scaling [25]) for high dimensional latent variables, those for which $k$ is on the same order as $n$. However, the approximation is not critical to any of our arguments, so even in this case we can use our framework to show that high dimensional latent variables can lead to Zipf's law; see Methods, Sec. M7.

At least in the simple case in which each element of $\mathbf{x}$ is independent and identically distributed conditioned on $z$, it is straightforward to show that the variance of the entropy is $\mathcal{O}(n^2)$. That's because the entropy is $n$ times the entropy of one element ($H_{\mathbf{x}|z}(z) = nH_{x_i|z}(z)$), so the variance of the total entropy is $n^2$ times the variance of the entropy of one element,

$$\mathrm{Var}_z\left[H_{\mathbf{x}|z}(z)\right] = n^2 \mathrm{Var}_z\left[H_{x_i|z}(z)\right], \tag{21}$$

which is $\mathcal{O}(n^2)$, and hence the variance of the energy is also $\mathcal{O}(n^2)$.

In the slightly more complex case in which each element of $\mathbf{x}$ is independent, but not identically distributed, conditioned on $z$, the total entropy is still the sum of the element-wise entropies: $H_{\mathbf{x}|z}(z) = \sum_i H_{x_i|z}(z)$. Now, though, each of the $H_{x_i|z}(z)$ can be different. In this case, for the variance to scale as $n^2$, the element-wise entropies must change in a correlated manner as $z$ varies, so that their changes reinforce, rather than cancel. Intuitively, the latent variable must control the entropy, such that for some settings of the latent variable the entropy of most of the elements is high, and for other settings the entropy of most of the elements is low.

For the completely general case, in which the elements of $x_i$ are not independent, essentially the same reasoning holds: the entropies of each element (suitably defined; see Methods, Sec. M6) must covary. Moreover, except in cases with highly precise cancellation, if the elementwise entropies $H_{x_i|z}(z)$ covary (with $\mathcal{O}(1)$, and positive, covariance), the variance of the total entropy scales as $n^2$ (see Methods, Sec. M6), and Zipf's law emerges. This result — that the variance of the energy scales as $n^2$ when the elementwise entropies covary — has been confirmed empirically for multi-neuron spiking data [17, 18] (though they did not assess Zipf's law).

We have shown that the variance of the energy is typically $\mathcal{O}(n^2)$. But is that broad enough to produce Zipf's law? The answer is yes, for the following reason. For Zipf's law to emerge, we need the distribution over energy to be broad over the whole range of ranks. For high-dimensional data, the number of possible observations, and hence the range of possible ranks, increases with $n$. In particular, the number of possible observations scales exponentially with $n$ (e.g. if each element of the observation is binary, the number of possible observations is $2^n$), so the logarithm of the number of observations, and hence the range of possible log-ranks, scales with $n$. Therefore, to obtain Zipf's law, the distribution over energy must be roughly constant over a region that scales with $n$. But that is exactly what latent variable models give us: the variance scales as $n^2$, so the width of the distribution is proportional to $n$, matching the range of log-ranks. Thus, the fact that the variance scales as $n^2$ means that Zipf's law is, very generically, likely to emerge for latent variable models in which the data is high dimensional.

We can, in fact, show that when the variance of the energy is $\mathcal{O}(n^2)$, Zipf's law is obeyed ever more closely as $n$ increases. Rewriting Eq. (8), but normalizing by $n$, we have

$$\frac{1}{n} \log r(\mathcal{E}) = \frac{\mathcal{E}}{n} + \frac{1}{n} \log P_S(\mathcal{E}). \tag{22}$$

The normalized log-rank and normalized energy now vary across an $\mathcal{O}(1)$ range, so if $\log P_S(\mathcal{E}) \sim \mathcal{O}(1)$, the last term will be small, and Zipf's law will emerge. If the variance of the energy is $\mathcal{O}(n^2)$, then $\log P_S(\mathcal{E})$ typically has this scaling. For example, consider a Gaussian distribution, for which $\log P_S(\mathcal{E}) \sim -(\mathcal{E} - \mathcal{E}_0)^2 / (2n^2)$. Because, as we have seen, the energy is proportional to $n$, the numerator and denominator both scale with $n^2$, giving $\log P_S(\mathcal{E})$ the required $\mathcal{O}(1)$ scaling. This argument is not specific to Gaussian distributions: if the variance of the energy is $\mathcal{O}(n^2)$, we expect $\log P_S(\mathcal{E})$ to display only $\mathcal{O}(1)$ changes as the energy changes by an $\mathcal{O}(n)$ amount.

This result turns out to be very robust. For instance, as we show in Methods, Sec. M8, even delta-function spikes in the distribution over energy (Fig. 5A) do not disrupt the emergence of Zipf's law as $n$ increases (Fig. 5B). (The distribution over energy is, of course, always a sum of delta-functions, as can be seen in Eq. (6). However, the delta-functions in Eq. (6) are typically very close together, and each one is weighted by a very small number, $e^{-\mathcal{E}}$. Here we are considering a delta-function with a large weight, as shown by the large spike in Fig. 5A.) However, "holes" in the probability distribution of the energy (i.e. regions of 0 probability, as in Fig. 5C) do disrupt the Zipf plot. That's because in regions where $P(\mathcal{E})$ is low, the energy decreases rapidly without the rank changing; this makes $\log P_S(\mathcal{E})$ very large and negative, disrupting Zipf's law (Fig. 5D). Between holes, however, we expect Zipf's law to be obeyed, as illustrated in Fig. 5D.

Importantly, we can now see why a model in which there is no latent variable, so the variance of the energy is $\mathcal{O}(n)$, does not give Zipf's law. (To see why the $\mathcal{O}(n)$ scaling of the variance is generic, see [20], and also Sec. 2.1.1 above.) In this case, the range of energies is $\mathcal{O}(\sqrt{n})$. This is much smaller than the $\mathcal{O}(n)$ range of the log ranks, and so Zipf's law will not emerge.

We have shown that high dimensional latent variable models lead to Zipf's law under two relatively mild conditions. First, the average entropy of each individual element of the data, $\mathbf{x}$, must covary as $z$ changes, and the average covariance must be $\mathcal{O}(1)$ (again, see Methods, Sec. M6, for the definition of elementwise entropy for non-independent models). Second, $P(\mathcal{E})$ cannot have holes; that is, it cannot have large regions where the probability approaches zero between regions of non-zero probability. Both conditions are typically satisfied for real world data.

### 2.5.1 Neural data

Neural data has been shown, in some cases, to obey Zipf's law [6, 7]. Here the data, which consists of spike trains from $n$ neurons, is converted to binary vectors, $\mathbf{x}(t) = (x_1(t), x_2(t), ...)$, with $x_i(t) = 1$ if neuron $i$ spiked in timestep $t$ and $x_i(t) = 0$ if there was no spike. The time index is then ignored, and the vectors are treated as independent draws from a probability distribution.

To model data of this type, we follow [7] and assume that each cell has its own probability of firing, which we denote $p_i(z)$. Here $z$, the latent variable, is the time since stimulus onset. This results in a model

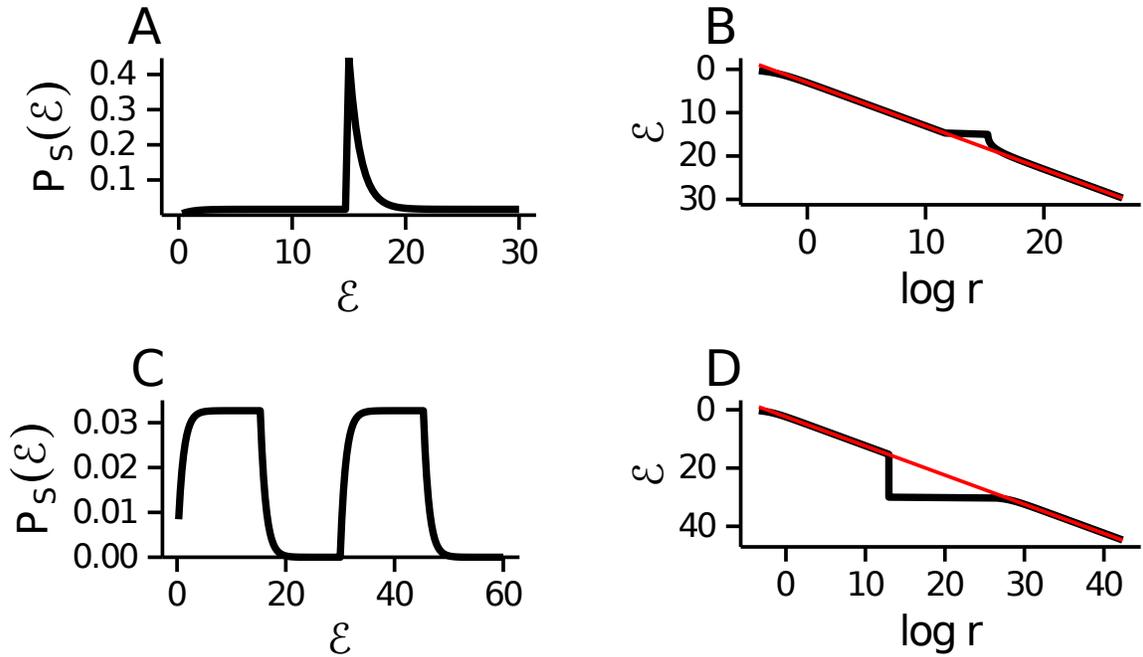

Figure 5: The relationship between $P_S(\mathcal{E})$ (left panel) and Zipf plots ($\mathcal{E}$ versus log-rank, right panel). As in Fig. 4, $\mathcal{E}$ increases downward on the $y$-axis in panels B and D. **A** and **B**. We bypassed an explicit latent variable model, and set $P(\mathcal{E}) = \text{Uniform}(\mathcal{E}; 0, 30)/2 + \delta(\mathcal{E} - 15)/2$. The deviation from Zipf's law, shown as a blip around $\mathcal{E} = 15$, is small. This is general: as we show in Methods, Sec. M8, departures from Zipf's law scale as $1/n$ even for large delta-function perturbations. **C** and **D**. We again bypassed an explicit latent variable model, and set $P(\mathcal{E}) = \text{Uniform}(\mathcal{E}; 0, 10)/2 + \text{Uniform}(\mathcal{E}; 20, 30)/2$. The resulting hole between $\mathcal{E} = 10$ and 20 causes a large deviation from Zipf's law.

in which the distribution over each element conditioned on the latent variable is given by

$$P(x_i|z) = p_i(z)^{x_i}\big(1 - p_i(z)\big)^{1-x_i} \tag{23}$$

The entropy of an individual element of **x** is therefore,

$$H_{x_i|z}(z) = -p_i(z)\log p_i(z) - \big(1 - p_i(z)\big)\log\big(1 - p_i(z)\big). \tag{24}$$

The entropy is high when $p_i(z)$ is close to $1/2$, and low when $p_i(z)$ is close to 0 or 1. Because time bins are typically sufficiently small that the probability of a spike is less than $1/2$, probability and entropy are positively correlated. Thus, if the latent variable (time since stimulus onset) strongly and coherently modulates most cells' firing probabilities — with high probabilities soon after stimulus onset (giving high entropy), and low probabilities long after stimulus onset (giving low entropy) — then the changes in entropy across different cells will reinforce, giving an $\mathcal{O}(n)$ change in entropy, and thus $\mathcal{O}(n^2)$ variance.

In our data, we do indeed see that firing rates are strongly and coherently modulated by the stimulus — firing rates are high just after stimulus onset, but they fall off as time goes by (Fig. 6A). Thus, when we combine data across all times, we see a broad distribution over energy (black line in Fig. 6B), and hence Zipf's law (black line in Fig. 6C). However, in any one time bin the firing rates do not vary much from one presentation of the stimulus to another, and so the energy distribution is relatively narrow (coloured lines in Fig. 6B). Consequently, Zipf's law is not obeyed (or at least is obeyed less strongly; coloured lines in Fig. 6C).

In our model of the neural data, Eq. (23), and in the neural data itself (Methods, Sec. M1), we assumed that the $x_i$ were independent conditioned on the latent variable. However, the independence assumption was not critical; it was made primarily to simplify the analysis. What is critical is that there is a latent variable that controls the population averaged firing rate, such that variations in the population averaged firing rate, and thus variations in the entropy, are $\mathcal{O}(1)$ — much larger than expected for neurons that are either independent or very weakly correlated. When that happens, the variance of the energy scales as $n^2$ (see Methods, Secs. M6 and M7).

This picture is consistent with a recent study by Tkačik and colleagues [17, 18]. They also recorded spike trains from retinal ganglion cells in response to time-varying visual stimulation, but much more of them than in our study (up to 120), and they analyzed simultaneously recorded, and thus correlated, spike trains. Although they did not construct Zipf plots, they did compute the variance of the energy. They found, as predicted by our analysis, that the variance of the energy was proportional to $n^2$. Moreover, the constant of proportionality depended on the visual stimulation: it was highest for natural scenes (movies consisting of swimming fish), second highest for time-varying full-field illumination (much like our stimulus), and lowest for random checkerboard. (See Methods, Sec. M2, for details.)

A plausible explanation for the $\mathcal{O}(n^2)$ scaling of the energy is that spatial correlations in the stimulus lead to correlations in individual firing rates, and thus to large fluctuations in the population averaged firing rate. This would suggest that the more correlated the stimuli are, the larger the fluctuations in the population averaged firing rate, and so the higher the variance of the energy. This is exactly what was seen in their data: the variance of the energy was a factor of 4-5 times larger for natural scenes and full field illumination, which exhibit long range spatial correlations, than for random checkerboard stimuli, which don't (see Methods, Sec. M2).

This discussion suggest that computing the variance of the energy, or making Zipf's plots, may not provide much information: $\mathcal{O}(n^2)$ variance and Zipf's law are relatively generic in neural data when there are large variations in the population averaged firing rate (the norm). However, it does suggest a way forward. Given that large variations in population-averaged firing rate lead to Zipf's law, it is sensible to ask what happens at fixed firing rate. For neural data, this is difficult to do experimentally, as it would require a very large number of trials with the same stimulus. Fortunately, Tkačik and colleagues have developed a very good model of the activity of large populations of neurons, at least in the retina — a maximum entropy model subject to constraints on the first two moments of the firing rates and the total number of spikes per bin

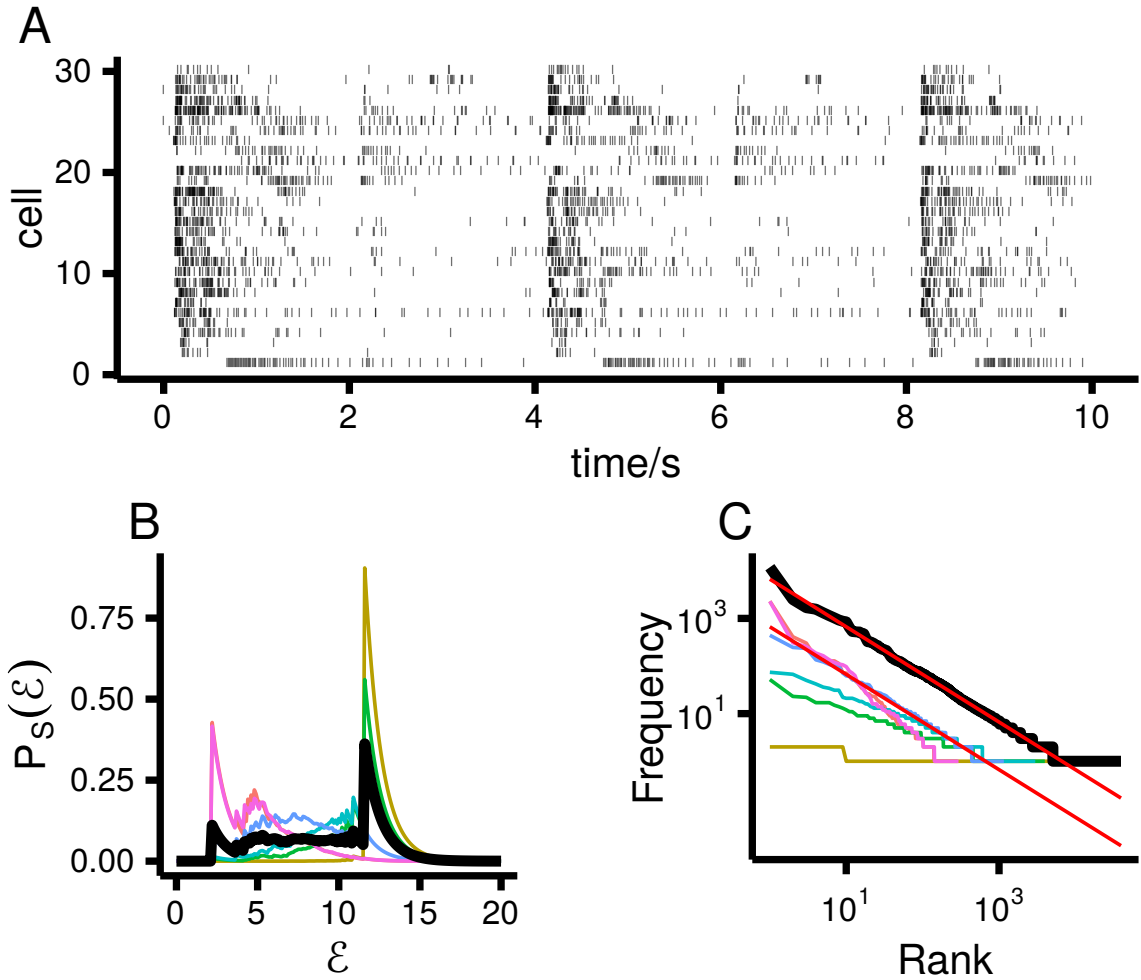

Figure 6: Neural data recorded from 30 mouse retinal ganglion cells stimulated by full-field illumination; see Methods, Sec. M1, for details. **A**. Spike trains from all 30 neurons. Note that the firing rates are strongly correlated across time. **B**. $P_S\left(\mathcal{E}|z\right)$ (coloured lines) when time relative to stimulus onset is the latent variable (see text and Methods, Sec. M1). The thick black line is $P_S\left(\mathcal{E}\right)$. **C**. Zipf plots for the data conditioned on time (coloured lines) and for all the data (black line). The red lines have slope $-1$.

[26]. Although not normally done, it can be cast as a latent variable model,

$$P(\mathbf{x}|\kappa) \propto \exp\left[\sum_i h_i x_i + \frac{1}{2n}\sum_{ij} x_i J_{ij} x_j\right] \Delta(\kappa - \bar{\nu}(\mathbf{x})) \qquad (25)$$

where $\bar{\nu}(\mathbf{x})$ is the population averaged firing rate,

$$\bar{\nu}(\mathbf{x}) \equiv \frac{1}{n}\sum_i x_i, \qquad (26)$$

$\kappa$ is the fraction of neurons that fire in a bin, and $\Delta(\cdot)$ is a slight generalization of the Kronecker delta: it is 1 if it's argument is 0 and 0 otherwise. Because of the Kronecker delta, which places a hard constraint on the fraction of spikes per bin, this model is hard to analyze. However, as we show in Methods, Sec. M9, in the large $n$ limit it is well approximated by a simpler model, one in which the Kronecker delta is replaced by an "external field" whose strength, $h(\kappa)$, is a function of $\kappa$,

$$P_h(\mathbf{x}|\kappa) \propto \exp\left[\sum_i \bigl(h(\kappa) + h_i\bigr)x_i + \frac{1}{2n}\sum_{ij} x_i J_{ij} x_j\right]. \qquad (27)$$

(The subscript "$h$" in Eq. (27) is to remind us that we have replaced the Kronecker delta in Eq. (25) with an external field.) The external field, $h(\kappa)$, is chosen so that the mean value of the population averaged firing rate is equal to $\kappa$,

$$\sum_{\mathbf{x}} P_h(\mathbf{x}|\kappa)\bar{\nu}(\mathbf{x}) = \kappa. \qquad (28)$$

If the variance of $\bar{\nu}(\mathbf{x})$ scales as $1/n$ in the large $n$ limit, which is the typical case, then $P_h(\mathbf{x}|\kappa)$ approaches $P(\mathbf{x}|\kappa)$. It's easy to see why: if the variance of $\bar{\nu}(\mathbf{x})$ goes to zero, then as a function of $\kappa$, $P_h(\mathbf{x}|\kappa)$ becomes proportional to a delta function centered at $\kappa = \bar{\nu}(\mathbf{x})$, yielding the distribution in Eq. (25). (See Methods, Sec. M9, for additional details.)

The distribution $P_h(\mathbf{x}|\kappa)$ is in a relatively standard form, and so we know that it is unlikely to exhibit Zipf's law. It would, however, be interesting to explore its behavior numerically, using values of $h_i$ and $J_{ij}$ derived from experimental measurements. There may, for instance, be values of $h(\kappa)$ (and thus values of the population averaged firing rate) for which the distribution over energy is especially broad, suggesting that those firing rates are in some sense special. In addition, it would be interesting to compute PEEV — the proportion of the energy variance explained by the number of spikes per bin — for this model. If it approaches 1 in the large $n$ limit, then firing rate accounts for most of the energy variance; if it doesn't approach 1, then it is likely that there is at least one other relevant latent variable.

## 2.6 Exponential family latent variable models

Recently, Schwab *et al.* [19] showed that a relatively broad class of models for high-dimensional data, a generalization of a so-called superstatistical latent variable model [27],

$$P(\mathbf{x}|\mathbf{g}) \propto \exp\left[-n\sum_{\mu=1}^m g_\mu O_\mu(\mathbf{x})\right], \qquad (29)$$

can give rise to Zipf's law. Importantly, in Schwab's model, when they refer to "latent variables," they are not referring to our fully general latent variable (which we call $z$) but to $g_\mu$, the natural parameters of an exponential family distribution. To make this explicit, and also to make contact with our model, we rewrite Eq. (29) as

$$P(\mathbf{x}|z) \propto \exp\left[-n\sum_{\mu=1}^m g_\mu(z) O_\mu(\mathbf{x})\right] \qquad (30)$$

where the dimensionality of $z$ can be lower than $m$. (See Methods, Sec. M10.1 for the link between Eqs. (29) and (30).)

If $m$ were allowed to be arbitrarily large, Eq. (30) could describe any distribution $P(\mathbf{x}|z)$. However, under Schwab *et al.*'s model $m$ can't be arbitrarily large; it must be much less than $n$ (as we show explicitly in Methods, Sec. M10.2). This puts several restrictions on Schwab *et al.*'s model class. In particular, it does not include many flexible models that have been fit to data. A simple example is our model of neural data (Eq. (23)). Writing this distribution in exponential family form gives

$$P(\mathbf{x}|z) \propto \exp\left[-n \sum_{\mu=1}^{n} \log\left(p_\mu(z)^{-1} - 1\right)(x_\mu/n)\right]. \tag{31}$$

Even though there is only one "real" latent variable, $z$ (the time since stimulus onset), there are $n$ natural parameters, $g_\mu = \log\left(p_\mu(z)^{-1} - 1\right)$. Consequently, this distribution falls outside of Schwab *et al.*'s model class. This is but one example; more generally, any distribution with $n$ natural parameters $g_\mu(z)$ falls outside of Schwab *et al.*'s model class whenever the $g_\mu(z)$ have a nontrivial dependence on $\mu$ and $z$ (as they did in Eq. (31)). This includes models in which sequence length is the latent variable, as these models require a large number of natural parameters (something that is not immediately obvious; see Sec M10.3).

The restriction to a small number of natural parameters also rules out high dimensional latent variable models — models in which the number of latent variable is on the order of $n$. That's because such models would require at least $\mathcal{O}(n)$ natural parameters, much more than are allowed in Schwab *et al.*'s model. Although we have so far restricted our analysis to low dimensional latent variable models, our framework can easily handle high dimensional ones. In fact, the restriction to low dimensional latent variables was needed only to approximate the mean energy by the entropy. That approximation, however, was not necessary; we can instead reason directly: as long as changes in the latent variable (now a high dimensional vector) lead to $\mathcal{O}(n)$ changes in the mean energy — more specifically, as long as the variance of the mean energy with respect to the latent variable is $\mathcal{O}(n^2)$ — Zipf's law will emerge. Alternatively, whenever we can reduce a model with a high dimensional latent variable to a model with a low dimensional latent variable, we can use the framework we developed for low dimensional latent variables (see Methods, Sec. M7). The same reduction cannot be carried out on Schwab *et al.*'s model, as in general that will take it out of the exponential family with a small number of natural parameters (see Methods, Sec. M10.2).

Besides the restrictions associated with a small number of natural parameters, there are two further restrictions; both prevent Schwab *et al.*'s model from applying to word frequencies. First, the observations must be high-dimensional vectors. However, words have no real notion of dimension. In contrast, our theory is applicable even in cases for which there is no notion of dimension (here we are referring to the theory in Sec. 2.2.1; the additional theory presented in Sec. 2.5 is applicable only when the data is indeed high-dimensional). Second, the latent variable must be continuous, or sufficiently dense that it can be treated as continuous. However, the latent variable for words is categorical, with a fixed, small number of categories (the part-of-speech).

Finally, our analysis makes it is relatively easy to identify scenarios in which Zipf's law does not emerge, something that can be hard to do under Schwab *et al.*'s framework. Consider, for example, the following model of data consisting of $n$-dimensional binary vectors,

$$P(\mathbf{x}|z) \propto \exp\left[-h \sum_{i} x_i + A\cos z \sum_{i} x_i \cos\theta_i + A\sin z \sum_{i} x_i \sin\theta_i\right] \tag{32}$$

where $\theta_i \equiv 2\pi i/n$, $h$ and $A$ are constant, and $z$ ranges from 0 to $2\pi$. Although this is in Schwab *et al.*'s model class, it does not display Zipf's law. To see why, note that it can be written

$$P(\mathbf{x}|z) \propto \exp\left[-h \sum_{i} x_i + A \sum_{i} \cos\left(z - \theta_i\right) x_i\right]. \tag{33}$$

This is a model of place fields on a ring: the activity of neuron $i$ is largest when its preferred orientation, $\theta_i$, is equal to $z$, and smallest when its preferred orientation is $z + \pi$. Because of the high symmetry of the model, the entropy is almost independent of $z$. In particular, changes in $z$ produce $\mathcal{O}(1)$ variations in the entropy (see Methods, Sec. M10.4); much smaller than the $\mathcal{O}(n)$ variations needed to produce Zipf's law.

This example suggests that any model in which changes in the latent variable cause uniform translation of place fields, without changing there height or shape, should not display Zipf's law. And indeed, non-Zipfian behavior was found in a numerical study of Gaussian place fields in one dimension [18]. Note, though, that if the amplitude of the place fields ($A$ in our model) or the overall firing rate ($h$ in our model) depends on a latent variable, then the population would exhibit Zipf's law. These conclusions emerge easily from our framework, but are harder to extract from that of Schwab *et al.*

In conclusion, while Schwab *et al.*'s approach is extremely valuable, it does have some constraints. We were able to relax those constraints, and thus show that latent variables induce Zipf's law in a wide array of practically relevant cases (word frequencies, data with variable sequence length, and simultaneously recorded neural data). Notably, all of these lie outside the class that Schwab *et al.*'s approach can handle. In addition, our analysis allowed us to easily identify scenarios in which the latent variable model lies in Schwab *et al.*'s model class, but Zipf's law does not emerge.

## 3 Discussion

We have shown that it is possible to understand, and explain, Zipf's law in a variety of domains. Our explanation consists of two parts. First, we derived an exact relationship between the shape of a distribution over log frequencies (energies) and Zipf's law. In particular, we showed that the broader the distribution, the closer the data comes to obeying Zipf's law. This was an extension of previous work showing that if a dataset has a broad, and perfectly flat, distribution over log frequencies (e.g. if a random draw gives very common elements, like "a" and rare elements, like "frequencies" the same proportion of the time), then Zipf's law must emerge [6]. Importantly, our extension allowed us to reason about how deviations from a perfectly flat distribution over energy manifest in Zipf plots. Second, we showed that if there is a latent variable that controls the typical frequency of observations, then mixing together different settings of the latent variable gives a broad range of frequencies, and hence Zipf's law. This is true even if the distributions over frequency conditioned on the latent variable are very narrow. Thus, Zipf's law can emerge when we mix together multiple non-Zipfian distributions. This is important because non-Zipfian distributions are the typical case (see Sec. 2.1.1), and are thus easy to understand.

When Zipf's law is observed, it is an empirical question whether or not it is due to our mechanism. Motivated by this observation, we derive a measure (percentage of explained variance, or PEEV) that allows us to separate out, and account for, the contribution of different latent variables to the observation of Zipf's law. We found that our mechanism was indeed operative in three domains: word frequencies, data with variable sequence length, and neural data. We were also able to show that while variable sequence length can give rise to Zipf's law on it's own, it was not the primary cause of Zipf's law in an antibody sequence dataset.

For words, the latent variable is the part of speech. As described in Sec. 2.3, parts of speech with a grammatical function (e.g. conjunctions, like "a") have a few, common words, whereas parts of speech that denote something in the world (e.g. nouns, like "frequencies") have many, rare words. Varying the latent variable therefore induces a broad range of characteristic energies (or frequencies), giving rise to Zipf's law.

For data with variable sequence length, the latent variable is the sequence length. There are many possible long sequences, so each long sequence is rare (high-energy). In contrast, there are few possible short sequences, so each short sequence is common (low-energy). Mixing across short and long sequences, and everything in between, gives a broad range of energies, and hence Zipf's law. We examined the role of sequence length in two datasets: randomly generated words and antibody sequences, both of which display Zipf's law [5, 12]. For the former, randomly generated words, sequence length was wholly responsible for Zipf's law. For the latter, antibody sequences, it formed only a small contribution. We were able to make these assessments quantitative, by computing the percentage of explained variance, or PEEV. And indeed, a recent model by Desponds *et al.* indicates that for antibodies, Zipf's law at each sequence length is most likely due to random growth and decay processes [23].

For high-dimensional data, small changes in the energy (or entropy) of each element of the observation can reinforce to give a large change overall, and hence Zipf's law. As an example, we considered multi-neuron spiking data, for which the latent variable is the time since stimulus onset. Just after stimulus onset, the firing rate of almost every cell (and hence the energy associated with those cells), is elevated. In contrast, long after stimulus onset, the firing rate of almost every cell (and hence the energy associated with those cells) is lower. As all the cells' energies change in the same direction (high just after stimulus onset, and low long after stimulus onset), the changes reinforce, and so produce $\mathcal{O}(n)$ changes in the total energy. Consequently, whenever the population firing rate varies with time, Zipf's law will almost always appear. This is true regardless of what is causing the variation: it could be a stimulus, or it could be low dimensional internal network dynamics. Thus, our framework is consistent with the recent observation that in salamander retina the variance of the energy scales as $n^2$ (the scaling needed for Zip's law to emerge), with higher variance when the stimulus induces larger covariation in the firing rates [17, 18]. This does not, of course, imply that the retina implements an uninteresting transformation from stimulus to neural response. However, our findings do have implications for the interpretation of observations of Zipf's law.

Our work shows that there are two types of datasets in which we expect Zipf's law to emerge generically. First, for the reason mentioned above, any dataset in which the sequence length varies (and is thus a latent variable) will display Zipf's law if the distribution over sequence length is sufficiently broad. Second, any high-dimensional dataset will display Zipf's law if the entropy of each element of the observation changes with the latent variable, and if those changes are correlated.

Previous authors have pointed out that latent variables models have interesting properties when the data is high-dimensional. As discussed in Sec. 2.6, Schwab *et al.* [19] were the first to show that a relatively broad class of latent variable models describing high-dimensional data give rise to Zipf's law. Their result, however, carries some restrictions: it applies only to exponential family distributions with continuous latent variables and a small number of natural parameters. Taking a completely different approach, we were able to demonstrate a more general result: we showed that Zipf's law emerges for discrete and high-dimensional latent variables (it can emerge even when the latent variable is the same dimension as the data), for exponential family distributions with a large number of natural parameters, and for models in which the data is a sequence of varying length. Importantly, none of of the datasets that we considered lie within the class considered by Schwab *et al.* [19]. However, the fact that Schwab *et al.*'s analysis applies to a restricted class of models should not detract from its importance: they were the first that we know of to show that Zipf's law could arise without fine tuning.

In addition, in work that anticipated some forms of latent variable models, Macke and colleagues examined models with common input [28], similar to the model in Eq. (23), as well as simple feedforward spiking neuron models [29]. They showed that both exhibit diverging heat capacity, for which the variance of the energy is $\mathcal{O}(n^2)$. Although they did not explicitly explore the connection to Zipf's law, in the latter study [29] they noted that the diverging heat capacity should lead to Zipf's law.

These findings have important implications in fields as diverse as biology and linguistics. In biology, one explanation for Zipf's law is that biological systems sit at a special thermodynamic state, the critical point [6, 15–18]. However, our findings indicate that Zipf's law emerges from phenomena much more familiar to biologists: unobserved states that influence the observed data. In fact, as mentioned above, for neural data our analysis shows that Zipf's law will emerge whenever the average firing rate in a population of neurons varies over time. Such time variation is common in neural systems, and can be due to external stimuli, low dimensional internal dynamics, or both.

For words, we showed that individual parts of speech do not obey Zipf's law; it is only by mixing together different parts of speech with different characteristic frequencies that Zipf's law emerges. This has an important consequence for other explanations of Zipf's law in language. In particular, the observation that individual parts of speech do not obey Zipf's law is inconsistent with any explanation of Zipf's law that fails to distinguish between parts of speech [2, 9–12, 30].

In all of these domains, the observation of Zipf's law is important because it may point to the existence of some latent variable structure. It is that structure, not Zipf's law itself, that is likely to provide insight into statistical regularities in the world.

# Methods

## M1  Neural data

The neural data in Fig. 6 was acquired by electrophysiological recordings of 3 isolated mouse retinas, yielding 30 ganglion cells. The recordings were performed on a multielectrode array using the procedure described in [31, 32]. Full field flashes were presented on a Sony LCD computer monitor, delivering intermittent flashes (2 s of light followed by 2 s of dark, repeated 30 times) of white light to the retina [33]. All procedures were performed under the regulation of the Institutional Animal Care and Use Committee of Weill Cornell Medical College (protocol #0807-769A) and in accordance with NIH guidelines.

Spikes were binned at 20 ms, and $x_i$ was set to 1 if cell $i$ spiked in a bin and zero otherwise. To give us enough samples to plot Zipf's law, we estimated $p_i(z)$, the probability that neuron $i$ spikes in bin $z$, from data using the model in Eq. (23), and drew $10^6$ samples from that model. To construct the distributions of energy conditioned on the latent variable — the coloured lines in Figs. 6B and C — we treated samples that occurred within 100 ms as if they had the same latent variable (so, for example, $P_S(\mathcal{E}|z=300)$ is shorthand for the smoothed distribution over energy for spike trains in the five bins between 300 and 400 ms). Finally, to reduce clutter, we plotted lines only for $z=0$ ms, $z=300$ ms etc.

## M2  The variance of the energy in large scale neuronal recordings

Here we estimate the scaling of the variance of the energy in retinal recordings performed by Tkačik and colleagues [17, 18]. Figures 4B and S4A-B from reference [18] show the variance of the energy divided by $n$, $\text{Var}_\mathbf{x}[\mathcal{E}(\mathbf{x})]/n$, versus an artificial temperature, $T$, for various populations sizes, $n$. (Note that in the plots in reference [18], the $y$-axis appears to be mislabeled: according to Eq. (S21) from that reference, the $y$-axis is the variance of the energy divided by $n^2$, whereas it is actually the variance of the energy divided by $n$.) In Fig. 7 below we replot their data at $T=1$ (the observed temperature, and thus the observed probability distribution). As can be seen in that figure, the variance of the energy normalized to $n$ is well described by the relationship

$$\frac{\text{Var}_\mathbf{x}[\mathcal{E}(\mathbf{x})]}{n} \approx a + bn. \qquad (34)$$

Nonzero $b$ indicates that the variance of the energy scales as $n^2$, for which we expect Zipf's law to emerge. And indeed, for three different "movies" (natural scenes, time-varying full-field illumination, and a spatially and temporally modulated random checkerboard), $b$ was nonzero. The movie that produced the highest $b$ (around 0.0052) was the one with natural scenes; the movie with the next highest $b$ (around 0.0041) was time-varying full-field illumination (much like our stimulus); and the movie with the lowest $b$ (around 0.0010) was the random checkerboard.

The values of $b$ in Fig. 7 may seem small (the largest is 0.0052). However, the raw numbers are not very meaningful. To put the slopes in perspective, consider a simple model in which the distribution over energy is flat between 0 and $n\epsilon_0$, and after that falls off sharply. For such a model, the variance of the energy is $n^2 \epsilon_0^2/12$; this corresponds to $b = \epsilon_0^2/12$. Under the approximation that the energy is close to the entropy, the maximum value of $\epsilon_0$ is the entropy (in nats) of a Bernoulli variable with probability $1/2$, which is $\log 2$. Thus, $\epsilon_0$ normalized to its maximum is given by

$$\frac{\epsilon_0}{\log 2} = \frac{(12b)^{1/2}}{\log 2}. \qquad (35)$$

For $b = 0.0052$ (natural movies), 0.0041 (full field illumination), and 0.0010 (random checkerboard), $\epsilon_0$ is about 30%, 27% and 13% of its maximum value, respectively.

## M3  PEEV, and the law of total variance

The law of total variance [24] is well known in statistics; it decomposes the total variance into the sum of two terms. Here we briefly review this law in the context of latent variable models, and then discuss how it is related to PEEV.

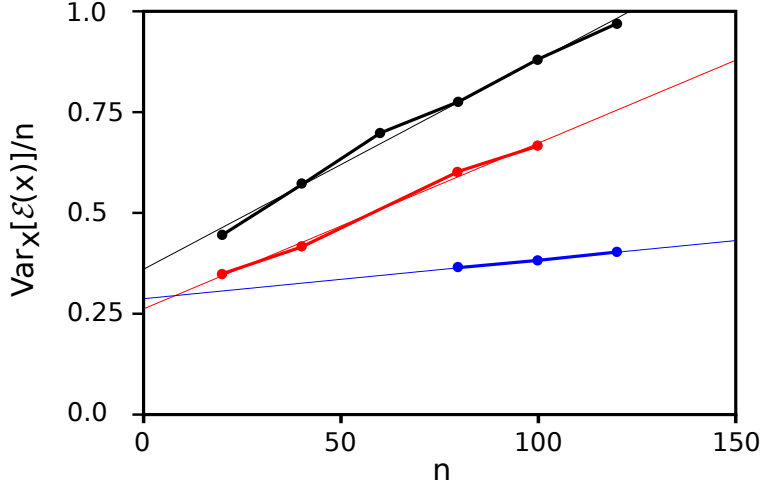

Figure 7: The variance of the energy normalized to $n$ for population activity from retinal ganglion cells responding to three different movies. Dots are data (which are connected by thick lines); thin lines are least squares fits. Black: natural scenes. Red: full field illumination. Blue: random checkerboard. The data was extracted from the plots in reference [18]: natural scenes (black) came from Fig. 4B; full field illumination (red) came from Fig. S4B, and the random checkerboard (blue) came from Fig. S4A. We used $T = 1$ (corresponding to points along the red vertical line in each of tho plots in reference [18]).

The energy, $\mathcal{E}(x)$, can be trivially decomposed as

$$\mathcal{E}(x) = \mathrm{E}_{x|z}\left[\mathcal{E}(x)\right] + \left(\mathcal{E}(x) - \mathrm{E}_{x|z}\left[\mathcal{E}(x)\right]\right) \tag{36}$$

where the first term, $\mathrm{E}_{x|z}\left[\mathcal{E}(x)\right]$, is the mean energy conditioned on $z$,

$$\mathrm{E}_{x|z}\left[\mathcal{E}(x)\right] = \int \mathcal{E}(x) P(x|z)\, dx. \tag{37}$$

The two terms in Eq. (36), $\mathrm{E}_{x|z}\left[\mathcal{E}(x)\right]$ and $\left(\mathcal{E}(x) - \mathrm{E}_{x|z}\left[\mathcal{E}(x)\right]\right)$, are uncorrelated, so the variance of $\mathcal{E}(x)$ is the sum of their variances,

$$\mathrm{Var}_x\left[\mathcal{E}(x)\right] = \mathrm{Var}_z\left[\mathrm{E}_{x|z}\left[\mathcal{E}(x)\right]\right] + \mathrm{Var}_{z,x}\left[\mathcal{E}(x) - \mathrm{E}_{x|z}\left[\mathcal{E}(x)\right]\right], \tag{38}$$

where $\mathrm{Var}_x\left[...\right]$ is the variance with respect to $P(x)$ and $\mathrm{Var}_{x,z}\left[...\right]$ is the variance with respect to $P(x,z)$. As is straightforward to show, the second term can be rearranged to give the law of total variance,

$$\mathrm{Var}_x\left[\mathcal{E}(x)\right] = \mathrm{Var}_z\left[\mathrm{E}_{x|z}\left[\mathcal{E}(x)\right]\right] + \mathrm{E}_z\left[\mathrm{Var}_{x|z}\left[\mathcal{E}(x)\right]\right]. \tag{39}$$

This is the same as Eq. (17) of the main text, except here we use $x$ rather than $\mathbf{x}$.

We can identify two contributions to the variance. The first, $\mathrm{Var}_z\left[\mathrm{E}_{x|z}\left[\mathcal{E}(x)\right]\right]$, is the variance of the expected energy, $\mathrm{E}_{x|z}\left[\mathcal{E}(x)\right]$, induced by changes in the latent variable, $z$. This represents the contribution to the total energy variance from the latent variable (i.e. the contribution from changes in the peak of $P(\mathcal{E}|z)$ as $z$ changes) and, under our mechanism, is the contribution that gives rise to Zipf's law. The second, $\mathrm{E}_z\left[\mathrm{Var}_{x|z}\left[\mathcal{E}(x)\right]\right]$, is the variance of the energy, $\mathrm{Var}_{x|z}\left[\mathcal{E}(x)\right]$, for a fixed setting of the latent variable, averaged over the latent variable, $z$. This represents the contribution from the width of $P(\mathcal{E}|z)$. The proportion of explained energy variance (PEEV) — that is, the portion explained by the first contribution — is the ratio of the first quantity to the total variance of the energy,

$$\mathrm{PEEV} \equiv \frac{\mathrm{Var}_z\left[\mathrm{E}_{x|z}\left[\mathcal{E}(x)\right]\right]}{\mathrm{Var}_x\left[\mathcal{E}(x)\right]}. \tag{40}$$

This quantity ranges from 0, indicating that $z$ explains none of the energy variance, to 1, indicating that $z$ explains all of the energy variance. PEEV therefore describes how much the latent variable contributes to the observation of Zipf's law, though it should be remembered that PEEV may be large even if the total energy variance is narrow, and hence Zipf's law is not obeyed.

### M3.1  Computing PEEV

To compute PEEV, we need to estimate, from data, the distribution over energy given the latent variable, and the distribution over the latent variable. Here we consider the case in which the latent variable is category, and each observation, $x$, falls into a single, known, category. In more realistic cases, $P(z|x)$ must be estimated from a model and $P(x)$ from data, from which $P(x|z)$ and $P(z)$ can be obtained using Bayes' theorem.

The starting point is the number of observations, and the category, of each possible value of $x$. For instance, for words, we took a list of words, their frequencies, and their parts of speech from [21]. We then used the frequencies to estimate the probability of each observation, and, finally, turned those into an energy via Eq. (3): $\mathcal{E}(x) = -\log P(x)$. The empirical distribution over energy, $P(\mathcal{E})$, and over energy given the latent variable, $P(\mathcal{E}|z)$, was therefore a set of delta functions, with each delta-function weighted by the probability of its corresponding observation,

$$P(\mathcal{E}) = \sum_x P(x)\,\delta\bigl(\mathcal{E} - \mathcal{E}(x)\bigr), \tag{41a}$$

$$P(\mathcal{E}|z) = \sum_x P(x|z)\,\delta\bigl(\mathcal{E} - \mathcal{E}(x)\bigr). \tag{41b}$$

The first expression is the same as Eq. (6); it is repeated here for convenience.

To compute the terms relevant to PEEV (Eq. (40)), we need moments of both the total energy and the energy conditioned on $z$. These are given, respectively, by

$$\mathrm{E}_x\bigl[\mathcal{E}^k(x)\bigr] = \sum_x P(x)\,\mathcal{E}^k(x), \tag{42a}$$

$$\mathrm{E}_{x|z}\bigl[\mathcal{E}^k(x)\bigr] = \sum_x P(x|z)\,\mathcal{E}^k(x). \tag{42b}$$

Then, to compute the variances required for PEEV, we use

$$\mathrm{Var}_{x|z}[\mathcal{E}(x)] = \mathrm{E}_{x|z}\bigl[\mathcal{E}^2(x)\bigr] - \bigl(\mathrm{E}_{x|z}[\mathcal{E}(x)]\bigr)^2, \tag{43a}$$

$$\mathrm{Var}_z\bigl[\mathrm{E}_{x|z}[\mathcal{E}(x)]\bigr] = \mathrm{E}_z\Bigl[\bigl(\mathrm{E}_{x|z}[\mathcal{E}(x)]\bigr)^2\Bigr] - \bigl(\mathrm{E}_z\bigl[\mathrm{E}_{x|z}[\mathcal{E}(x)]\bigr]\bigr)^2, \tag{43b}$$

where

$$\mathrm{E}_z\bigl[\mathrm{E}_{x|z}[\mathcal{E}^k(x)]\bigr] = \mathrm{E}_x\bigl[\mathcal{E}^k(x)\bigr], \tag{44a}$$

$$\mathrm{E}_z\Bigl[\mathrm{E}_{x|z}[\mathcal{E}(x)]^k\Bigr] = \sum_z P(z)\,\bigl(\mathrm{E}_{x|z}[\mathcal{E}(x)]\bigr)^k. \tag{44b}$$

## M4  $\mathrm{Var}[\log P(z)]$ is $\mathcal{O}\bigl((\log\mathrm{Var}[z])^2\bigr)$

To compute the variance of the energy for variable length data, we stated that the variance of $\log P(z)$ is small compared to the variance of $z$. Here we first show that for Li's model [12], the variance of $\log P(z)$ is $\mathcal{O}(1)$; we then show that in general the variance of $\log P(z)$ is at most $\mathcal{O}\bigl((\log\mathrm{Var}[z])^2\bigr)$.

For Li's model, the probability of observing a sequence of length $z$ is proportional to the probability of drawing $z$ letters followed by a blank. For an alphabet with $M$ letters, this is given by

$$P(z) = \frac{1}{M}\left(\frac{M}{M+1}\right)^z. \tag{45}$$

The leading factor of $1/M$ ensures that the distribution is properly normalized (note that $z$ ranges from 1 to $\infty$). Given this distribution, it is straightforward to show that

$$\mathrm{Var}_z\left[\log P(z)\right] = M(M+1)\left(\log\left[1+\frac{1}{M}\right]\right)^2. \tag{46}$$

Using the fact that $\log(1+\epsilon) \leq \epsilon$, we see that the right hand side is bounded by $(M+1)/M$. Thus, for Li's model, $\mathrm{Var}_z\left[\log P(z)\right]$ is indeed $\mathcal{O}(1)$.

To understand how the variance of $\log P(z)$ scales in general, we note that the variance is bounded by the second moment,

$$\mathrm{Var}_z\left[\log P(z)\right] = \sum_z P(z)\left[\log P(z)\right]^2 - \left(\sum_z P(z)\log P(z)\right)^2 \leq \sum_z P(z)\left[\log P(z)\right]^2. \tag{47}$$

Shortly we'll maximize the second moment with the variance of $z$ fixed. When we do that, we find that the second moment is small compared to $\sigma_z^2$, the variance of $z$. However, the analysis is somewhat complicated, so first we provide the intuition.

The main idea is to note that for unimodal distributions, the number of sequence lengths with appreciable probability is proportional to the standard deviation of $z$. If we make the (rather crude) approximation that $P(z)$ is nonzero only for $n_0$ sequence lengths, where $n_0 \propto \sigma_z$, then the right hand side of Eq. (47) is maximum when $P(z) = 1/n_0$, and the corresponding value is $(\log n_0)^2$. Consequently, the second moment of $\log P(z)$ is at most $\mathcal{O}\bigl((\log \sigma_z)^2\bigr)$, giving us the very approximate bound

$$\mathrm{Var}_z\left[\log P(z)\right] \leq \mathcal{O}\left(\frac{\log \sigma_z^2}{2}\right)^2 \tag{48}$$

where we used $\log \sigma_z = (1/2)\log \sigma_z^2$.

This does indeed turn out to be correct, at least in the limit that $\sigma_z^2$ is sufficiently large. To show this rigorously, we take the usual approach: we use Lagrange multipliers to maximize the second moment of $\log P(z)$ with constraints on the total probability and the variance. This gives us

$$\frac{\partial}{\partial P(z)}\left[\sum_{z'}P(z')\left(\log P(z')\right)^2 - (\gamma^2+\alpha^2-1)\left(\sum_z P(z')-1\right) - \frac{\gamma^2 Z^2}{e^2}\left(\sum_{z'}P(z')\,z'^2 - \mu^2 - \sigma_z^2\right)\right] = 0 \tag{49}$$

where $\mu$ is the mean value of $z$,

$$\mu \equiv \int dz\, P(z)\, z. \tag{50}$$

We use $\gamma^2+\alpha^2-1$ and $\gamma^2 Z^2/e^2$ as our Lagrange multiplier to simplify later expressions. As is straightforward to show (taking into account the fact that $\mu$ depends on $P(z)$), Eq. (49) is satisfied when $P(z)$ is given by

$$P(z) = \exp\left[-1 - \left(\gamma^2 + \alpha^2 - \frac{\gamma^2 Z^2 \mu^2}{e^2} + \frac{\gamma^2 Z^2(z-\mu)^2}{e^2}\right)^{1/2}\right]. \tag{51}$$

The parameters $\gamma$, $\alpha$ and $Z$ must be chosen so that $P(z)$ is normalized to 1 and has variance $\sigma_z^2$. However, because $z$ is a positive integer, finding these parameters analytically is, as far as we know, not possible. We can, though, make two approximations that ultimately do yield analytic expressions. The first is to allow $z$ to be continuous. This turns sums (which are needed to compute moments) into integrals, and results in an error in those sums that scales as $1/\sigma_z$. That error is negligible in the limit that $\sigma_z$ is large (the limit of interest here). The second is to allow $z$ to be negative. This will increase the maximum second moment of $\log P(z)$ at fixed $\sigma_z^2$ (because we are expanding the space of probability distributions), and so result in a slightly looser bound. But the bound will be sufficiently tight for our purposes.

The problem of choosing the parameters $\gamma$, $\alpha$ and $Z$ is now much simpler, as we can do integrals rather than sums. We proceed in three steps: first, we show that none of the relevant moments depend on $\mu$, so we set it to zero and at the same time eliminate $\alpha$; second, we use the fact that $P(z)$ must be properly normalized to express $Z$ in terms of $\gamma$; and third, we explicitly compute the second moment of $\log P(z)$ and the variance of $\sigma_z^2$.

To see that the second moment of $P(z)$ and the variance of $z$ do not depend on $\mu$, make the change of variables $z = z' + \mu$ and let $\alpha = \gamma^2 Z^2 \mu^2 / e^2$. That yields a distribution $P(z')$ that is independent of $\mu$. Thus, $\mu$ does not effect either the second moment of $\log P(z)$ or the variance of $z$, and so without loss of generality we can set both $\mu$ and $\alpha$ to zero. We thus have

$$P(z) = \exp\left[-1 - \gamma\left(1 + Z^2 z^2/e^2\right)^{1/2}\right]. \tag{52}$$

It is convenient to make the change of variables $z = ye/Z$, yielding

$$P(y) = \frac{e^{-\gamma(1+y^2)^{1/2}}}{Z(\gamma)} \tag{53}$$

where $Z$, which now depends on $\gamma$ to ensure that $P(z)$ (and thus $P(y)$) is properly normalized, is given by

$$Z(\gamma) = \int dy\, e^{-\gamma(1+y^2)^{1/2}}. \tag{54}$$

In terms of $P(y)$, the two quantities of interest are

$$\mathrm{E}_z\left[(\log P(z))^2\right] = \mathrm{E}_y\left[\left(1 + \gamma(1+y^2)^{1/2}\right)^2\right] \tag{55a}$$

$$\sigma_z^2 = \frac{e^2}{Z^2(\gamma)} \mathrm{E}_y\left[y^2\right]. \tag{55b}$$

These expectations can be expressed as modified Bessel functions of the second kind (as can be seen by making the change of variables $y = \sinh\theta$). However, the resulting expressions are not very useful, so we don't reproduce them here. There are, though, two easy limits: large and small $\gamma$. In the large $\gamma$ limit, $P(y)$ is Gaussian, yielding

$$\lim_{\gamma\to\infty} \mathrm{E}_z\left[(\log P(z))^2\right] = (\gamma + 3/2)^2 + \mathcal{O}(1) \tag{56a}$$

$$\lim_{\gamma\to\infty} \sigma_z^2 = \frac{e^{2(\gamma+1)}}{2\pi}\left(1 + \mathcal{O}(1/\gamma)\right). \tag{56b}$$

And in the small $\gamma$ limit, $P(y)$ is Laplacian, and we have

$$\lim_{\gamma\to 0} \mathrm{E}_z\left[(\log P(z))^2\right] = 5 + \mathcal{O}(\gamma) \tag{57a}$$

$$\lim_{\gamma\to 0} \sigma_z^2 = \frac{e^2}{2} + \mathcal{O}(\gamma). \tag{57b}$$

As is straightforward to show, in both limits the second moment of $\log P(z)$ obeys the inequality

$$\mathrm{E}_z\left[(\log P(z))^2\right] \le \left(c_0 + \frac{\log\sigma_z^2}{2}\right)^2 \tag{58}$$

where

$$c_0 = \frac{\sqrt{20} - \log(e^2/2)}{2} \approx 1.58. \tag{59}$$

We verified numerically that the inequality in Eq. (58) is satisfied over the whole range of $\gamma$, from 0 to $\infty$. Thus, although very naive arguments were used to derive the bound given in Eq. (48), it is substantially correct.

## M5 Models in which the latent variable is the sequence length

For models in which the sequence length is the latent variable, Zipf's law is obeyed if the energy is proportional to the sequence length, $z$; that is, if the energy is $\mathcal{O}(z)$. To determine whether this scaling holds, we start with Eq. (13) of the main text, which tells us that when the latent variable is sequence length, the total distribution is a simple function of the latent variables: $P(\mathbf{x}) = P(\mathbf{x}|z) P(z)$ where $z$ is the dimension of $\mathbf{x}$ (the sequence length). Thus, the energy is given by

$$\mathcal{E}(\mathbf{x}) = \sum_{i=1}^{z} \mathcal{E}_i(\mathbf{x}) - \log P(z). \tag{60}$$

where

$$\mathcal{E}_i(\mathbf{x}) \equiv -\log P(x_i | x_{i-1} \ldots x_1). \tag{61}$$

Assuming the value of $x_i$ isn't perfectly determined by the values of $x_1, ..., x_{i-1}$ (the typical case), each term in the sum over $z$ is $\mathcal{O}(1)$, and so the first term in Eq. (60) is $\mathcal{O}(z)$. As we saw in the previous section, the variance of $\log P(z)$ is small compared to the variance of $z$. Consequently, the energy is $\mathcal{O}(z)$.

## M6 Latent variable models with high dimensional non-conditionally independent data

In the main text we argued that for a conditionally independent model — a model in which each element of $\mathbf{x}$ is independent conditioned on $z$ — the variance of the entropy typically scales as $n^2$. Extending this argument to complex joint distributions is straightforward, and, in fact, follows closely the method used in Sec. M5.

The first step is to note that, just as in the conditionally independent case, $\log P(\mathbf{x}|z)$ can be written as a sum over each element of $x_i$,

$$\log P(\mathbf{x}|z) = \sum_i \log P(x_i | z, x_1, x_2, ..., x_{i-1}). \tag{62}$$

Taking the expectation with respect to $P(\mathbf{x}|z)$ (and negating) gives the entropy, which consists of a sum of $n$ terms,

$$H_{\mathbf{x}|z}(z) = \sum_{i=1}^{n} h_i(z) \tag{63}$$

where $h_i(z)$ is the entropy of $P(x_i|z, x_1, x_2, ..., x_{i-1})$, averaged over $x_1$ to $x_{i-1}$, with $z$ fixed,

$$h_i(z) \equiv \mathrm{E}_{\mathbf{x}|z} \left[ -\log P(x_i|z, x_1, x_2, ..., x_{i-1}) \right]. \tag{64}$$

The variance of the entropy is thus given by

$$\mathrm{Var}_z \left[ H_{\mathbf{x}|z}(z) \right] = \sum_{ij} \mathrm{Cov}_z \left[ h_i(z), h_j(z) \right]. \tag{65}$$

Just as in the main text, if the individual entropies (the $h_i$) have, on average, $\mathcal{O}(1)$ covariance as $z$ changes, then the variance of the entropy is $\mathcal{O}(n^2)$. This illuminates a special case in which we do not see Zipf's law: if the $x_1, x_2, ..., x_{i-1}$ determine the value of $x_i$ when $i > i_0$ (independent of $n$), then the entropy, $h_i$, is zero whenever $i > i_0$. If this were to happen, the variance of the entropy would scale at most as $i_0^2$, independent of $n$; far smaller than the required $\mathcal{O}(n^2)$ scaling. However, for most types of data, including neural data, each neuron has considerable independent noise (due, for instance, to synaptic failures [34]), so the $h_i$ typically remain finite for all $i$.

For complex joint distribution, the $h_i$ can be hard to reason about and/or compute. However, here we argue that it is possible to reason about the scaling of the covariance of the $h_i(z)$'s based on the scaling of the covariance of the elementwise entropies $H_{x_i|z}(z)$, which is often much easier to reason about. To see this, note that the $h_i$ can be written

$$h_i(z) = H_{x_i|z}(z) - I_i(z) \tag{66}$$

where, as in the main text, the first term is the elementwise entropy,

$$H_{x_i|z}(z) \equiv -\sum_{x_i} P(x_i|z) \log P(x_i|z), \tag{67}$$

and the second term is the mutual information between $x_i$ and $x_1$ to $x_{i-1}$, conditioned on z,

$$I_i(z) \equiv \mathrm{E}_{\mathbf{x}|z}\left[-\log \frac{P(x_i|z)}{P(x_i|z, x_1, x_2, ..., x_{i-1})}\right] = H_{x_i|z}(z) + \mathrm{E}_{\mathbf{x}|z}\left[\log P(x_i|z, x_1, x_2, ..., x_{i-1})\right]. \tag{68}$$

Combining Eq. (66) with Eq. (63), we see that

$$\mathrm{Var}_z\left[H_{\mathbf{x}|z}(z)\right] = \sum_{ij} \mathrm{Cov}_z\left[H_{x_i|z}(z), H_{x_j|z}(z)\right] - \sum_{ij} 2\mathrm{Cov}_z\left[H_{x_i|z}(z), I_j(z)\right] + \sum_{ij} \mathrm{Cov}_z\left[I_i(z), I_j(z)\right]. \tag{69}$$

If the $H_{x_i|z}(z)$ covary, then the first term is $\mathcal{O}(n^2)$. In this situation it would require very precise cancellation for the whole expression to be $\mathcal{O}(n)$. Such cancellation could occur if, for instance, $H_{x_i|z}(z) = I_i(z) + \text{const}$. However, unless the constant were zero, so $x_{i-1}...x_1$ determine the value of $x_i$, it is unclear how this could occur. Thus, as claimed in the main text, except in cases in which there is highly precise cancellation, if the elementwise entropies $H_{x_i|z}(z)$ covary (with $\mathcal{O}(1)$ covariance), the variance of the total entropy will scale as $n^2$.

## M7 High dimensional latent variables

So far we have restricted our analysis to low dimensional latent variables. However, this is not absolutely necessary, and in fact high dimensional latent variable can induce Zipf's law the same way low dimensional ones can: if different settings of the latent variable result in $\mathcal{O}(n)$ differences in the mean energy, Zipf's law will emerge. The main difference in the analysis is that we can no longer approximate the mean energy by the entropy, as we did in Eq. (20). However, it is not actually necessary to make this approximation; it is merely convenient, as it allows us to work with the entropy, an intuitive, well-understood quantity. Indeed, if we work directly with the mean energy, Eq. (18), we can see that covariation in the individual energies leads to Zipf's law — just as the covariation in the individual entropies led to Zipf's law in the previous section.

To show this explicitly, we break Eq. (18) into one term for each element of $\mathbf{x}$,

$$\mathrm{E}_{\mathbf{x}|z}\left[-\log P(\mathbf{x})\right] = \sum_{\mathbf{x}} l_i(\mathbf{x}) \tag{70}$$

where

$$l_i(\mathbf{x}) \equiv \mathrm{E}_{\mathbf{x}|z}\left[-\log P(x_i|x_1, x_2, ..., x_{i-1})\right]. \tag{71}$$

Then, writing the variance of the mean energy in terms of the $l_i$, we have

$$\mathrm{Var}_z\left[\mathrm{E}_{\mathbf{x}|z}\left[\log P(\mathbf{x})\right]\right] = \sum_{ij} \mathrm{Cov}_z\left[l_i, l_j\right]. \tag{72}$$

If the $l_i$ have $\mathcal{O}(1)$ covariance, the variance of the energy is $\mathcal{O}(n^2)$, and Zipf's law emerges. The intuition is that each element of $\mathbf{x}$ contributes to the energy, $-\log P(\mathbf{x})$. These contributions (or their expected values) change with the latent variable, and if they all change in the same direction, then the overall change in the energy is $\mathcal{O}(n)$, so the variance is $\mathcal{O}(n^2)$.

While the above analysis provides the underlying intuition, in practical situations the $l_i$ may be difficult to compute. We therefore provide an alternative approach. For definiteness, we'll set the dimension of the latent variable to the dimension of the data, $n$; to make this explicit, we'll replace $z$ by $\mathbf{z}$ ($\equiv z_1, z_2, ..., z_n$). In addition, we'll assume, without loss of generality, that each latent variable — each $z_i$ — has an $\mathcal{O}(1)$ range. We'll also assume that each latent variable has an $\mathcal{O}(1)$ effect on the mean energy; this ensures that the average energy has sensible scaling with $n$.

Because each of the latent variables has a small effect, they need to act together to produce the $\mathcal{O}(n)$ variability in the mean energy that is required for Zipf's law. Specifically, if any two latent variables, say $z_i$ and $z_j$, have the same effect on the average energy (either both increasing it or both decreasing it), they need to be positively correlated; if they have the opposite effect (one increasing it and the other decreasing it), they need to be negatively correlated. When this doesn't hold — when correlations are essentially arbitrary, or non-existent — variations in $\mathbf{z}$ have an $\mathcal{O}(\sqrt{n})$ effect on the average energy. In this regime, the variance of the average energy is $\mathcal{O}(n)$, and Zipf's law does not emerge. We thus conclude, at least tentatively (and perhaps not surprisingly) that the $z_i$ must to be correlated for Zipf's law to emerge.

To see this more quantitatively, we make a first-order Taylor series expansion of the expected energy,

$$\mathrm{E}_{\mathbf{x}|\mathbf{z}}\left[\mathcal{E}(\mathbf{x})\right] \approx \mathrm{E}_{\mathbf{x}|\mathbf{z}=\boldsymbol{\mu}}\left[\mathcal{E}(\mathbf{x})\right] + \sum_{i=1}^{n}(z_i - \mu_i) \left. \frac{\partial \mathrm{E}_{\mathbf{x}|\mathbf{z}}\left[\mathcal{E}(\mathbf{x})\right]}{\partial z_i} \right|_{\mathbf{z}=\boldsymbol{\mu}}. \tag{73}$$

Because each of the $z_i$ has an $\mathcal{O}(1)$ range and an $\mathcal{O}(1)$ effect on the mean energy, each term in the sum is $\mathcal{O}(1)$. Thus, if the higher order terms in Eq. (73) can be neglected, the $z_i$ have to be correlated for the variance of the average energy to scale as $\mathcal{O}(n^2)$; if they are not correlated, the variance is $\mathcal{O}(n)$.

Of course, ignoring higher order terms in high dimensions is dangerous, as the number of terms grows rapidly with $n$ (the number of $k^{\text{th}}$ order terms is proportional to $n^k$). However, it turns out to give the right intuition: the Efron-Stein inequality [35–37], along with the assumption that each latent variable has an $\mathcal{O}(1)$ effect on the energy, ensures that if the $z_i$ are independent, the variance of the energy is indeed $\mathcal{O}(n)$. Thus, a necessary condition for Zipf's law to emerge is that the $z_i$ are correlated, as has been pointed out previously [18] (in Supporting Information).

The fact that correlations are necessary to produce Zipf's law provides a natural approach to understanding models with high dimensional latent variables. The approach relies on the observation that sufficiently correlated variables have a "long" direction — a direction along which the typical size of $|\mathbf{z}|$ is $\mathcal{O}(n)$ (rather than $\mathcal{O}(\sqrt{n})$, as it is for uncorrelated latent variables). We can, therefore, construct a low dimensional latent variable that measures distance along that direction, and then use the analysis developed above for low dimensional latent variables.

Here we illustrate this idea for binary variables, $x_i = 0$ or 1. For definiteness, and because it makes the ideas more intuitively accessible, we consider a concrete setting: neural data, with as many latent variables as neurons. As in the main text, $x_i = 1$ corresponds to one or more spikes in a small time bin and $x_i = 0$ corresponds to no spikes. Because the long direction in latent variable space depends on the distribution $P(\mathbf{z})$, it would seem difficult to make general statements. However, in this example the data comes from neural spike trains, and so we can make use of the fact that firing rates of neurons often covary. Thus, a very natural low dimensional latent variable, which we denote $\nu$, is the population averaged firing rate,

$$\nu = \frac{1}{n} \sum_i p_i(\mathbf{z}) \tag{74}$$

where $p_i(\mathbf{z})$ is the probability that $x_i = 1$ given $\mathbf{z}$,

$$p_i(\mathbf{z}) = \mathrm{E}_{\mathbf{x}|\mathbf{z}}\left[x_i\right] = \sum_{\mathbf{x}} x_i P(\mathbf{x}|\mathbf{z}). \tag{75}$$

For this model the element-wise entropies have a very simple form,

$$H_{x_i|\mathbf{z}}(\mathbf{z}) = -p_i(\mathbf{z}) \log p_i(\mathbf{z}) - \big(1 - p_i(\mathbf{z})\big) \log \big(1 - p_i(\mathbf{z})\big). \tag{76}$$

We'll assume that all the $p_i(\mathbf{z})$ are less than 1/2, something that is satisfied for realistic spike trains if the time bins aren't too large. Consequently, increasing $p_i(\mathbf{z})$ increases the element-wise entropy of neuron $i$.

We need two conditions for Zipf's law to emerge: the variance of $\nu$ must be $\mathcal{O}(1)$, and $\mathcal{O}(1)$ changes in $\nu$ must lead to $\mathcal{O}(1)$, and positively correlated, changes in the element-wise entropies (assuming, as discussed in the previous section, there isn't very precise cancellation). So long as the firing rates go up and down together, both conditions are satisfied, and Zipf's law emerges. If, on the other hand, the firing rates are not positively correlated on average, the variance of $\nu$ is $\mathcal{O}(1/\sqrt{n})$, and the population averaged firing rate provides no information about Zipf's law. This is an important example, as the population averaged firing rate is easy to estimate from data.

In summary, high dimensional latent variables are, from a conceptual point of view, no different than low dimensional ones: both lead to Zipf's law if different settings of the latent variables lead to average energies that differ by $\mathcal{O}(n)$. However, in the high dimensional case, each latent variable has a small effect on the energy, so a necessary condition for Zipf's law to emerge is that the latent variables are correlated. This turns out to be helpful: the correlations can lead naturally to a low dimensional latent variable, for which our analysis of low dimensional latent variables applies.

## M8 Peaks in $P(\mathcal{E})$ do not disrupt Zipf's law

In the main text, we noted that while holes in the distribution over energy, $P(\mathcal{E})$, disrupt Zipf's law, peaks in this distribution do not. To see this explicitly, take an extreme case: $P(\mathcal{E})$ is composed of a delta function at $\mathcal{E} = \mathcal{E}_0$, weighted by $\alpha$, combined with a smooth component, $f(\mathcal{E})$, that integrates to $1 - \alpha$. Here $\alpha$ may be any number between 0 and 1, and in particular it need not be exponentially small in the energy, as it is in Eq. (6). For this case, we can compute $P_S(\mathcal{E})$ explicitly using Eq. (9),

$$\frac{1}{n} \log P_S(\mathcal{E}) = \frac{1}{n} \log \left[\alpha e^{-(\mathcal{E} - \mathcal{E}_0)} \Theta(\mathcal{E} - \mathcal{E}_0) + f_S(\mathcal{E})\right] \tag{77}$$

where $f_S$ is $f$ smoothed by an exponential kernel, $\Theta$ is the Heaviside step function, and we have normalized by $n$ to give us the quantity relevant for determining the size of departures from Zipf's law (see Eq. (22)). The term $e^{-(\mathcal{E} - \mathcal{E}_0)} \Theta(\mathcal{E} - \mathcal{E}_0)$ ranges from 0 to 1, so $\log P_S(\mathcal{E})$ can be bounded above and below,

$$\frac{1}{n} \log \big(f_s(\mathcal{E})\big) \leq \frac{1}{n} \log P_S(\mathcal{E}) \leq \frac{1}{n} \log \big(\alpha + f_s(\mathcal{E})\big). \tag{78}$$

Assuming the distribution $f_s(\mathcal{E})$ is such that the first term vanishes in the large $n$ limit (so that without the delta function Zipf's law would hold), then the last term must also vanish in the large $n$ limit. Thus, even delta-function singularities do not prevent convergence to Zipf's law, so long as they occur on top of a finite baseline.

## M9 Constraints on spike count can be modeled as an external field

As discussed in the main text, a very good model of the activity of large populations of neurons, at least in the retina, is a maximum entropy model with constraints on the first two moments of the firing rate and on the total number of spikes per bin [26]. The latent variable version of that model, given in Eq. (25) in the main text, is hard to analyze because of the Kronecker delta. For instance, even when the $J_{ij}$ are zero, it is not, as far as we know, possible to compute the partition function exactly unless all the $h_i$ are identical. Here, though, we show that it can be converted to the much simpler model given in Eq. (27), in which the hard constraint on the number of spikes is replaced by an external field.

For generality, we consider a conditional distribution with an arbitrary energy, rather than the specific one in Eq. (25),

$$P(\mathbf{x}|\kappa) = \frac{e^{-E(\mathbf{x})} \Delta\big(\kappa - \overline{\nu}(\mathbf{x})\big)}{\sum_{\mathbf{x}'} e^{-E(\mathbf{x}')} \Delta\big(\kappa - \overline{\nu}(\mathbf{x}')\big)} \tag{79}$$

where, as in Eq. (26), $\overline{\nu}(\mathbf{x}) = \sum_i x_i/n$ is the population averaged firing rate, and, recall, $\Delta(\cdot)$ is a slight generalization of a Kronecker delta: it's 1 if its argument is 0 and 0 otherwise.

Consider an alternative distribution, which will turn out to closely approximate $P(\mathbf{x}|\kappa)$,

$$P_h(\mathbf{x}|\kappa) = \frac{e^{-E(\mathbf{x})+nh(\kappa)\overline{\nu}(\mathbf{x})}}{\sum_{\mathbf{x}'} e^{-E(\mathbf{x}')+nh(\kappa)\overline{\nu}(\mathbf{x}')}} \tag{80}$$

where $h(\kappa)$ is chosen so that the mean value of $\overline{\nu}(\mathbf{x})$ is equal to $\kappa/n$; it is defined implicitly from the equation

$$\frac{\sum_{\mathbf{x}} \overline{\nu}(\mathbf{x}) e^{-E(\mathbf{x})+nh(\kappa)\overline{\nu}(\mathbf{x})}}{\sum_{\mathbf{x}} e^{-E(\mathbf{x})+nh(\kappa)\overline{\nu}(\mathbf{x})}} = \kappa \,. \tag{81}$$

In essence, we are replacing the Kronecker delta in Eq. (79) with a local field, $h(\kappa)$, chosen so that $P_h(\mathbf{x}|\kappa)$ has the same mean as $P(\mathbf{x}|\kappa)$. If the variance of $\overline{\nu}(\mathbf{x})$ under $P_h(\mathbf{x}|\kappa)$ were small, then, intuitively at least, $P_h(\mathbf{x}|\kappa)$ would be close to $P(\mathbf{x}|\kappa)$. Away from critical points, we expect the variance of $\overline{\nu}(\mathbf{x})$ to be $\mathcal{O}(1/n)$. Even at a critical point, the variance of $\overline{\nu}(\mathbf{x})$ is typically order $1/n^\gamma$ with $\gamma > 0$ [20]. It is only if $P_h(\mathbf{x}|\kappa)$ displays Zipf's law that the variance of $\overline{\nu}(\mathbf{x})$ is $\mathcal{O}(1)$. If that's not the case — meaning, essentially, that $\kappa$ is the only relevant latent variable — then in the large $n$ limit the variance of $\overline{\nu}(\mathbf{x})$ vanishes, and $P_h(\mathbf{x}|\kappa)$ does indeed approach $P(\mathbf{x}|\kappa)$.

To make these ideas more rigorous, we note that $P_h(\mathbf{x}|\kappa)$ may be written

$$P_h(\mathbf{x}|\kappa) = \sum_{\kappa'} P(\mathbf{x}|\kappa') \, P(\kappa'|\kappa) \tag{82}$$

where

$$P(\kappa'|\kappa) = \frac{e^{h(\kappa)\kappa'} \sum_{\mathbf{x}} e^{-E(\mathbf{x})} \Delta\big(\kappa' - \overline{\nu}(\mathbf{x})\big)}{\sum_{\mathbf{x}} e^{-E(\mathbf{x})+nh(\kappa)\overline{\nu}(\mathbf{x})}} \,. \tag{83}$$

Direct substitution verifies that the right hand side of Eq. (82) really is $P_h(\mathbf{x}|k)$. To show that $P_h(\mathbf{x}|\kappa)$ is close to $P(\mathbf{x}|\kappa)$, all we need to do is show that $P(\kappa'|\kappa)$ is very sharply peaked around $\kappa' = \kappa$. Our starting point is to note that for any function $f(\kappa)$,

$$\sum_{\kappa'} P(\kappa'|\kappa) \, f(\kappa') = \frac{\sum_{\mathbf{x}} e^{-E(\mathbf{x})+nh(\kappa)\overline{\nu}(\mathbf{x})} f\big(\overline{\nu}(\mathbf{x})\big)}{\sum_{\mathbf{x}'} e^{-E(\mathbf{x}')+nh(\kappa)\overline{\nu}(\mathbf{x}')}} \,. \tag{84}$$

Combining this with Eq. (81), we see that the mean value of $\kappa'$ is $\kappa$. The variance of $\kappa'$ is

$$\text{Var}_{P(\kappa'|\kappa)}[\kappa'] = \text{Var}_{P_h(\mathbf{x}|\kappa)}[\overline{\nu}(\mathbf{x})] \,. \tag{85}$$

As discussed above, so long as $P_h(\mathbf{x}|\kappa 0$ doesn't display Zipf's law, the variance of $\overline{\nu}(\mathbf{x})$ should vanish in the large $n$ limit, implying that $P(\kappa'|\kappa)$ does indeed have a narrow peak around $\kappa' = \kappa$.

At finite, but large, $n$, we can make use of the narrow peak to achieve a very accurate total distribution. Let us write the total distribution under $P_h(\mathbf{x}|\kappa)$, denoted $P_h(\mathbf{x})$, as

$$P_h(\mathbf{x}) = \sum_{\kappa} P_h(\mathbf{x}|\kappa) P_h(\kappa) \tag{86}$$

where $P_h(\kappa)$ may be different from the true prior, $P(\kappa)$. Using Eq. (82), this becomes

$$P_h(\mathbf{x}) = \sum_\kappa P(\mathbf{x}|\kappa) \sum_{\kappa'} P(\kappa|\kappa') P_h(\kappa'). \tag{87}$$

If $P_h(\kappa')$ is chosen to satisfy

$$\sum_{\kappa'} P(\kappa|\kappa') P_h(\kappa') = P(\kappa), \tag{88}$$

then $P_h(\mathbf{x})$ will be exactly equal to $P(\mathbf{x})$. In fact, Eq. (88) typically can't be perfectly satisfied, but if $P(\kappa)$ is relatively smooth on a scale over which $P(\kappa|\kappa')$ varies, it can be almost satisfied. Thus, $P_h(\mathbf{x}|\kappa)$ can be a good model of $P(\mathbf{x}|\kappa)$ even when $n$ is finite.

Finally, we note that when

$$E(\mathbf{x}) = -\sum_i h_i x_i - \frac{1}{2n} \sum_{ij} x_i J_{ij} x_j, \tag{89}$$

we recover Eq. (27) of the main text.

## M10  Exponential family latent variable models: technical details

Schwab *et al.* [19] showed that Zipf's law emerges for a model in which the distribution over $\mathbf{x}$ given the latent variable is in the exponential family. By itself, the fact that the distribution is in the exponential family places no restrictions on the class of models. However, their derivation required other conditions to be satisfied, and those conditions do induce restrictions. In particular, their analysis does not apply to models with a large number of natural parameters (it thus does not apply when the latent variable is high dimensional), models in which the latent variable is discrete, and models in which the latent variable is the dimension of the data. Here we show this explicitly.

### M10.1  The relationship between Schwab *et al.*'s model and our model

Schwab *et al.* formulated their model as a latent variable model conditioned on natural parameters, as written in the main text, Eq. (29). Hidden in Eq. (29) is the fact that the $g_\mu$ can be "tied": the parameters $g_\mu$ are drawn from a distribution that allows delta-functions, such as $\delta(g_1 - f(g_2))$ for some function $f$, or even $\delta(g_3 - g_3^*)$. To make this explicit, and to also make contact with our model, we rewrote Eq. (29) as a latent variable model conditioned on $z$ (Eq. (30)), where $z$ is a $k$-dimensional latent variable. Under this model it is easy to tie variables; for instance, letting $g_1 = z$ and $g_2 = f(z)$ (with $z$ one-dimensional) enforces the constraint $\delta(g_1 - f(g_2))$.

### M10.2  Number of latent variables

Here we show that the number of natural parameters ($m$ in Eqs. (29) and (30)) must be small compared to the dimension of the data, $n$. We start by sketching Schwab *et al.*'s [19] derivation, including many steps that were left to the reader in their paper. Their starting point is the expression for the energy of an observation,

$$-\log P(\mathbf{x}) = -\log \int dz\, P(z)\, e^{-n\mathbf{g}(z)\cdot \mathbf{O}(\mathbf{x}) - \log Z(z)}. \tag{90}$$

We have written the right hand side using the form given in Eq. (30), except that we explicitly include the partition function (Eq. (92) below), and we use dot products instead of sums. This integral is evaluated using the saddle-point method,

$$-\log P(\mathbf{x}) \approx n\mathbf{g}(z^*) \cdot \mathbf{O}(\mathbf{x}) + \log Z(z^*). \tag{91}$$

where $z^*$ maximizes the term in the exponent in Eq. (90). For the saddle point method to work — that is, for the above approximation to hold — the number of latent variables, $\dim(z)$, must be subextensive in $n$ (i.e. $\dim(z)/n \to 0$ as $n$ goes to infinity; see [38] for details).

The condition $\dim(z) \ll n$ does not place any restrictions on the number of natural parameters (the dimension of $\mathbf{g}$). But the next step in their derivation, computing the partition function (which is necessary for finding the energy of an observation), does. The log of the partition function is given by the usual expression,

$$\log Z(z) = \log \sum_{\mathbf{x}} e^{-n\mathbf{g}(z)\cdot\mathbf{O}(\mathbf{x})}. \tag{92}$$

In the large $n$ limit, the sum can be approximated as an integral over $\mathbf{O}$,

$$\log Z(z) = \log \int d\mathbf{O}\, e^{-n\mathbf{g}(z)\cdot\mathbf{O}+S(\mathbf{O})} \tag{93}$$

where $S(\mathbf{O})$ is the entropy at fixed $\mathbf{O}$,

$$e^{S(\mathbf{O})} = \sum_{x} \delta\left(\mathbf{O} - \mathbf{O}(\mathbf{x})\right). \tag{94}$$

Note that $\mathbf{O}$ is in fact a discrete variable. However, $e^{S(\mathbf{O})}$ becomes progressively denser as $n$ increases, and as $n \to \infty$, it becomes continuous. As with Eq. (90), the integral can be computed using the saddle point method, yielding

$$\log Z(z) \approx -n\mathbf{g}(z)\cdot\mathbf{O}^* + S(\mathbf{O}^*). \tag{95}$$

For this approximation to be valid, the dimension of $\mathbf{O}$, and hence the dimension of $\mathbf{g}$ (which is $m$), must be subextensive in $n$. Thus, Schwab *et al.*'s method applies to model in which $m \ll n$ (more technically, $m/n \to 0$ as $n \to \infty$). This restricts it to a relatively small number of natural parameters.

In sum, because Schwab *et al.*'s method involves an $m$-dimensional saddle-point integral over $\mathbf{O}$, it requires the dimensionality of $\mathbf{O}$ (and hence $\mathbf{g}$) to be small (i.e. $m/n \to 0$ as $n \to \infty$; again, see [38] for details). There are additional steps in their derivation. However, they are not trivial, and they do not lead to additional constraints on their model, so we do not reproduce them here.

Although high dimensional natural parameters are ruled out by Schwab *et al.*'s method, there are many interesting cases (e.g., models of neural data), in which the elements of $\mathbf{g}$ covary. In those cases, one might think that it would be possible to reduce a high-dimensional latent variable to a low-dimensional one, as we did in Sec. M7 above. While such a reduction is always possible, doing so typically takes the model out of Schwab *et al.*'s class. To see this in a simple setting, we reduce a model with one low-dimensional natural parameter, $g$, and one high-dimensional natural parameter, $\mathbf{g}$, to a model with just the low-dimensional natural parameter. (Here $g$ might represent the overall firing rate, and the other natural parameters, $\mathbf{g}$, might represent fluctuations around that rate.) The model is written

$$P(\mathbf{x}|g,\mathbf{g}) = e^{-gO(\mathbf{x})-\mathbf{g}\cdot\mathbf{O}(\mathbf{x})-\log Z(g,\mathbf{g})} \tag{96}$$

where $Z(g,\mathbf{g})$ is the partition function,

$$Z(g,\mathbf{g}) = \sum_{\mathbf{x}} e^{-gO(\mathbf{x})-\mathbf{g}\cdot\mathbf{O}(\mathbf{x})}. \tag{97}$$

Marginalizing over $\mathbf{g}$ yields

$$P(\mathbf{x}|g) = \int d\mathbf{g}\, e^{-gO(\mathbf{x})-\mathbf{g}\cdot\mathbf{O}(\mathbf{x})-\log Z(g,\mathbf{g})} P(\mathbf{g}|g) \equiv e^{-gO(\mathbf{x})-\psi(g,\mathbf{O}(\mathbf{x}))}. \tag{98}$$

The function $\psi(g, \mathbf{O}(\mathbf{x}))$ typically has an extremely complicated dependence on $g$ and $\mathbf{x}$. In fact, for all but the simplest model it is not even possible to calculate it analytically, as the partition function cannot be calculated analytically. Thus, $P(\mathbf{x}|g)$ can't be written in the exponential family with a single natural parameter. It can, of course, be written in the exponential family with an exponential number of natural parameters,

$$\psi(g, \mathbf{O}(\mathbf{x})) = \sum_{\mathbf{x}'} \psi(g, \mathbf{O}(\mathbf{x}'))\delta(\mathbf{x} - \mathbf{x}') \tag{99}$$

where $\delta(\mathbf{x} - \mathbf{x}')$ is the Kronecker delta, but this clearly takes it out of Schwab *et al.*'s model class. This is closely related to the fact that exponential family distributions are not closed under marginalization [39].

### M10.3 Latent variable is the sequence length

To show that a model with sequence length as the latent variable is outside of Schwab's class, we begin by writing the distribution in the form

$$P(\mathbf{x}|z) = \lim_{L \to \infty} e^{\log P(\mathbf{x}) - L\left(1 - \delta_{\dim(\mathbf{x}), z}\right)} \tag{100}$$

where $\delta_{ij}$ is the Kronecker delta ($\delta_{ij} = 1$ if $i = j$ and 0 otherwise) and, as above, $\dim(\cdot)$ denotes dimension (in this case the number of elements in $\mathbf{x}$). This distribution allows only values of $\mathbf{x}$ which have the correct length: if $\dim(\mathbf{x}) = z$, the second term in the exponent is zero, giving $P(\mathbf{x}|z) = P(\mathbf{x})$; in contrast, if $\dim(\mathbf{x}) \neq z$, the second term in the exponent is $-L$, giving a large negative contribution to the energy, and sending $P(\mathbf{x}|z \neq \dim(\mathbf{x})) \to 0$.

This distribution is not in the exponential family form, because the term $\delta_{\dim(\mathbf{x}), z}$ is not written as the product of a natural parameter (in this case a function of $z$) and a sufficient statistic (in this case a function of $\mathbf{x}$). It is not possible to write it as a single product, but it can be written as the sum of multiple products,

$$\delta_{\dim(\mathbf{x}), z} = \sum_i \delta_{z, i} \delta_{i, \dim(\mathbf{x})}. \tag{101}$$

This is now in the required form, because each term in the sum is the product of a natural parameter ($\delta_{z,i}$, which is function of $z$), and a sufficient statistic, ($\delta_{i, \dim(\mathbf{x})}$, which is a function of $\mathbf{x}$). Inserting this into Eq. (100) gives

$$P(\mathbf{x}|z) = \lim_{L \to \infty} e^{\log P(\mathbf{x}) - L\left(1 - \sum_i \delta_{z, i} \delta_{i, \dim(\mathbf{x})}\right)} \tag{102}$$

This is in the exponential family. However, there are $\mathcal{O}(n)$ terms in the sum, so it is not in Schwab *et al.*'s model class.

### M10.4 Entropy of a place field model

Here we compute the entropy, at fixed $z$, of the place field model in Eq. (33), and show that it depends very weakly on $z$. Because the distribution over $\mathbf{x}$ is conditionally independent given $z$, the entropy has a simple form,

$$H_{\mathbf{x}|z}(z) = \sum_i H_B\big(p(z - \theta_i)\big) \tag{103}$$

where $H_B(p)$ is the entropy (in nats) of a Bernoulli random variable,

$$H_B(p) \equiv -p \log p - (1-p) \log(1-p), \tag{104}$$

and $p(z - \theta_i)$ is the probability that $x_i = 1$ given $z$,

$$p(z - \theta_i) \equiv \frac{e^{-h + A\cos(z - \theta_i)}}{1 + e^{-h + A\cos(z - \theta_i)}}. \tag{105}$$

To understand how the entropy in Eq. (103) scales with $z$, we make the change of variables

$$z = \theta_j + \delta z \tag{106}$$

where $\theta_j$ is chosen to minimize $|\delta z|$. The mean value theorem tells us that for any smooth function $f(z)$,

$$f(z + \delta z) = f(z) + \delta z f'(z^*) \tag{107}$$

where prime denotes derivative and $z^*$ is between $z$ and $z + \delta z$. Consequently,

$$H_{\mathbf{x}|z}(z) = \sum_i H_B\big(p(\theta_j - \theta_i)\big) + \delta z \sum_i \frac{\partial H_B\big(p(z^* - \theta_i)\big)}{\partial z^*}. \tag{108}$$

Because the $\theta_i$ are evenly spaced, the first term is independent of $z$. Except at $p = 0$ or 1 (which are not allowed if $h$ and $A$ are finite), the sum over $i$ of the second term is $\mathcal{O}(n)$. The spacing between adjacent $\theta_i$ is $2\pi/n$, so $|\delta z| \leq \pi/n \sim \mathcal{O}(1/n)$. Consequently, the second term in Eq. (108) is $\mathcal{O}(1/n) \times \mathcal{O}(n) \sim \mathcal{O}(1)$, and changes in $z$ produce $\mathcal{O}(1)$ changes in the entropy.

## Acknowledgments

The authors would like to thank Yasser Roudi, John Hertz and David Schwab for valuable comments. LA and PEL were supported by the Gatsby Charitable Foundation; NC was supported by the NIH.